\documentclass{aa}  

\usepackage{graphicx}
\usepackage{txfonts}

\usepackage[]{hyperref}
\usepackage{natbib}
\usepackage{gensymb}

\begin{document}

   \title{On the relation between Ly$\alpha$ absorbers and local galaxy filaments}

   \subtitle{}

   \author{S. J. D. Bouma \and P. Richter \and M. Wendt}

   \institute{Institut f\"ur Physik und Astronomie, Universit\"at Potsdam, 
   Karl-Liebknecht-Str. 24/25, 14476 Golm, Germany\\
   \email{sbouma@astro.physik.uni-potsdam.de}}

   \date{Received XXX; accepted YYY}

% \abstract{}{}{}{}{} 
% 5 {} token are mandatory
 
  \abstract
   {The intergalactic medium (IGM) is believed to contain the majority of baryons in the universe 
   and to trace the same dark matter structure as galaxies, forming filaments and 
   sheets. Ly$\alpha$ absorbers, which sample the neutral component of the IGM, have been extensively 
   studied at low and high redshift, but the exact relation between Ly$\alpha$ absorption, galaxies 
   and the large-scale structure is observationally not well-constrained.}
   {In this study, we aim at characterising the relation between Ly$\alpha$ absorbers and nearby 
   overdense cosmological structures (galaxy filaments) at recession velocities $\Delta v \leq$ 6700~km\,s$^{-1}$ by using archival observational data from various instruments.}
   {We analyse 587 intervening Ly$\alpha$ absorbers in the spectra of 302 extragalactic 
   background sources obtained with the {\it Cosmic Origins Spectrograph} (COS) installed on 
   the Hubble Space Telescope (HST). We combine the absorption-line information with galaxy 
   data of five local galaxy filaments originally mapped by \citet{Courtois13}.}
   {Along the 91 sightlines that pass close to a filament, we identify 215 (227) Ly$\alpha$ 
   absorption systems (components). Among these, 74 Ly$\alpha$ systems are aligned in position and velocity 
   with the galaxy filaments, indicating that these absorbers and the galaxies trace the same 
   large-scale structure. The filament-aligned Ly$\alpha$ absorbers have a $\sim 90$ percent higher 
   rate of incidence ($d\mathcal{N}/dz=189$ for log $N$(H\,{\sc i}) $\geq 13.2$) and a mildly 
   shallower slope ($-\beta = -1.47$) of the column density distribution function than the general 
   Ly$\alpha$ population at $z=0$, reflecting the filaments' matter overdensity.       
   The strongest Ly$\alpha$ absorbers are preferentially found near galaxies or close to the axis 
   of a filament, although there is substantial scatter in this relation. 
   Our sample of absorbers clusters more strongly around filament axes than a randomly distributed 
   sample would do (as confirmed by a Kolmogorov-Smirnov test), but the clustering signal is 
   less pronounced than for the galaxies in the filaments.}
   {}


   \maketitle

%________________________________________________________________

%%%%%%%%%%%%%%%%%%%%%%%%%%%%%%%%%%%%%%%%%%%%%%%%%%%%%%%%%%%%%%%%%%%%%%%%

\section{Introduction}

It is generally accepted that the intergalactic medium (IGM) contains the majority of the baryons 
in the universe \citep[e.g.][]{Shull03,Lehner07,Danforth08,Richter08,Shull12,Danforth16}, 
making it a key component in understanding cosmological structure formation. It is estimated 
that about $30\,\%$ of the baryons at low-$z$ are in the form of photoionised hydrogen at a temperature of 
$\lesssim 10^4$~K \citep{Penton00,Lehner07,Danforth08,Shull12,Tilton12,Danforth16}, 
while the collapsed, shock-heated Warm-Hot Intergalactic Medium (WHIM) at T $\sim 10^5 - 10^6$~K 
contains at least $20\,\%$ \citep{Richter06,Richter06b,Lehner07}. 
Cosmological simulations indicate that the WHIM may contribute up to $50\,\%$ of the baryons in the 
low-$z$ universe \citep{Cen99,Dave99,Shull12,Martizzi19}.
This makes the IGM an important reservoir of baryons for the galaxies to fuel star formation. 
Indeed, to explain current rates of star formation, such a baryon reservoir is 
needed \citep[e.g.][]{Erb08,Prochaska09,Genzel10}. In this way, the evolution of galaxies is 
tied to the properties and the spatial distribution of the IGM.

The relation between the IGM and galaxies is not one-way, however, as galaxies influence their 
surroundings by ejecting hot gas into their circumgalactic medium (CGM) by AGN feedback 
\citep[e.g.][]{Bower06,Davies19} and, particularly in the early universe, by supernova explosions 
\citep{Madau01,Pallottini14}. Thus, there is a large-scale exchange of matter and energy
between the galaxies and the surrounding IGM and both environments influence the evolution of each 
other. One observational method to study the gas circulation processes between galaxies and
the IGM is the analysis of intervening Lyman $\alpha$ (Ly$\alpha$) absorption in the spectra
of distant Active Galactic Nuclei (AGN), which are believed to trace the large-scale gaseous 
environment of galaxies. Generally, an anticorrelation between Ly$\alpha$ absorber strength 
and galaxy impact parameter is found for absorbers relatively close to 
galaxies \citep[e.g.][]{Chen01,Bowen02,Wakker09,French17}.

The IGM is not only tied to the galaxies, but it is also expected to trace the dark matter 
distribution and can, therefore, give insights into the large-scale structure of the universe. 
This large-scale structure has been mapped by galaxy surveys like the 2-degree Field Galaxy Redshift 
Survey and the Sloan Digital Sky Survey (SDSS) \citep{Colless01,York00}. 
Studying the IGM absorber distribution at low redshift allows for a comparison with data from these 
galaxy surveys. Already more than 25 years ago, \citet{Morris93} studied the spectrum 
of the bright quasar 3C\,273 and mapped Ly$\alpha$ absorbers along this line of sight together 
with galaxies in its vicinity. They found that Ly$\alpha$ absorbers cluster less strongly around 
galaxies than the galaxies among themselves. This can be interpreted as most of the Ly$\alpha$ 
absorbers truly being intergalactic in nature, following the filamentary large-scale structure 
rather than the position of individual galaxies.

More recently, \citet{Tejos16} studied Ly$\alpha$ and O\,{\sc vi} absorption in a single sightline in 
regions between galaxy clusters. The detected overdensity of narrow and broad Ly$\alpha$ absorbers 
hints at the presence of filamentary gas connecting the clusters. A different approach was taken 
by \citet{Wakker15}. Instead of mapping gas along an isolated sightline, they used several sightlines 
passing through a known galaxy filament. By comparing the relation of Ly$\alpha$ equivalent width 
with both galaxy and filament impact parameters, \citet{Wakker15} conclude that Ly$\alpha$ absorbers 
are best described in the context of large-scale structure, instead of tracing individual galaxy haloes. 
While there is a relation between strong ($N$(H\,{\sc i}) $> 10^{15}$ cm$^{-2}$) 
absorbers and the CGM of galaxies, weak Ly$\alpha$ absorbers are more likely to be associated with 
filaments. This view is also supported by \citet{Penton02}, who find that weak absorbers 
do not show a correlation between equivalent width and impact parameter to the nearest galaxy, while 
stronger absorbers do. By comparing the position of their sample of Ly$\alpha$ absorbers relative to 
galaxies in filaments, they conclude that the absorbers align with the filamentary structure. Evidence for 
absorbers tracing an extensive, intra-group medium comes from other recent surveys of \citet{Stocke13}
and \citet{Keeney18}.

While the correlation between Ly$\alpha$ equivalent width and galaxy impact parameter seems to indicate 
that these absorbers {\it somehow} are associated with galaxies (e.g., by the gravitational potential), 
studies like \citet{Wakker15,Tejos16} show that at least some of the absorbers are associated with the
cosmological large-scale structure. Others studies \citep{Bowen02,Wakker09} conclude that their data 
simply does not yield any definite conclusions on this aspect \citep[see also][]{Penton02, Prochaska11,Tejos14}. 
Therefore, the question of how Ly$\alpha$ absorbers at $z=0$ are linked to galaxies and the large-scale 
cosmological structure is not yet resolved. Clearly, additional absorption-line studies that
improve the currently limited statistics on the absorber/galaxy connection are desired.

In this paper, we systematically investigate the properties of $z=0$ Ly$\alpha$ absorbers and their 
connection to the local galaxy environments and the surrounding large-scale structure. For this, 
we follow an approach similar to that of \citet{Wakker15}. We combine the information on local galaxy 
filaments mapped by \citet{Courtois13} with archival UV absorption line data from the 
Cosmic Origins Spectrograph (COS) installed on the \textit{Hubble Space Telescope} (\textit{HST}). 

Information on the galaxy sample used in this study is provided in Sect.~2. In Sect.~3, the HST/COS 
data are described and information on the absorption line measurements are given. Details on the 
galaxy filaments are presented in Sect.~4. In Sect.~5, we investigate the relation between absorbers and 
galaxies, whereas in Sect.~6 we focus on the relation between absorbers and filaments. In Sect.~7, we
discuss our findings and compare them with previous studies. Finally, we summarise and conclude 
our study in Sect.~8.

%__________________________________________________________________

\begin{figure}[bp]
\centering
\includegraphics[width=\hsize]{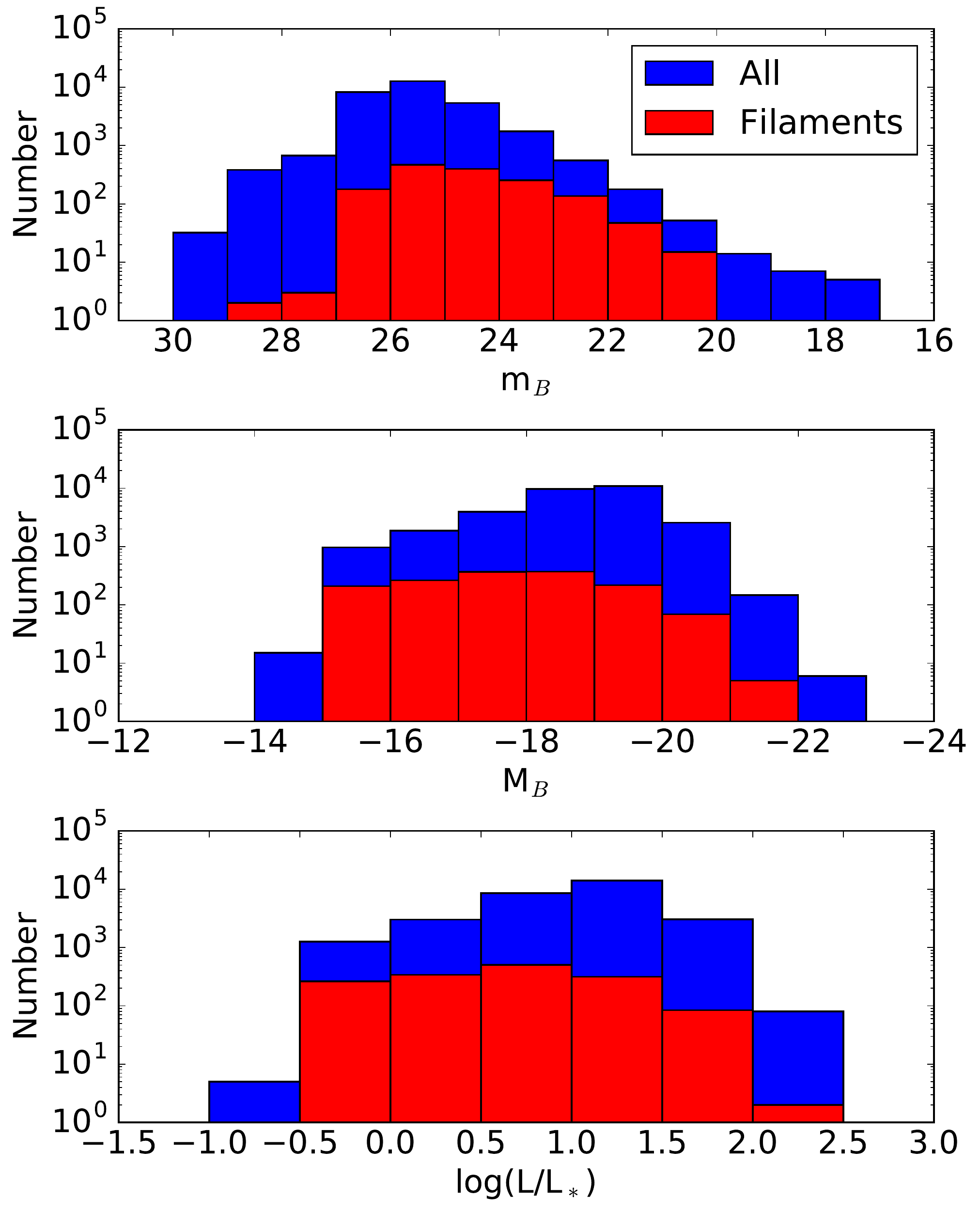}
\caption{
Histogram of apparent and absolute $B$-band magnitudes and luminosities for all galaxies of 
the V8k catalogue.
}
\label{histV8k}
\end{figure}

%__________________________________________________________________

%%%%%%%%%%%%%%%%%%%%%%%%%%%%%%%%%%%%%%%%%%%%%%%%%%%%%%%%%%%%%%%%%%%%%%%%

\section{Galaxy data}

\citet{Courtois13} used the V8k catalogue of galaxies to map galaxy filaments in the nearby universe. 
This catalogue is available from the Extragalactic Distance Database\footnote{\url{http://edd.ifa.hawaii.edu/}} \citep[EDD][]{Tully09}. 
It is a compilation of different 
surveys, including John Huchra's `ZCAT' and the IRAS Point Source Catalog redshift survey with its 
extensions to the Galactic plane \citep{Saunders00a,Saunders00b}. In total, the catalogue consists 
of $\sim$~30\,000 galaxies, all with velocities less than $8000$ km\,s$^{-1}$. It is complete up 
to $M_B$~=~$-16$ for galaxies at 1000~km\,s$^{-1}$, while at 8000~km\,s$^{-1}$, it contains 
one in 13 of the $M_B$~=~$-16$ galaxies. A radial velocity of 8000~km\,s$^{-1}$  corresponds 
to a cosmological distance of $d\sim(114$ km\,s$^{-1})h^{-1}_{70}$. 
The distance to the Centaurus Cluster ($v \sim 3000$~km\,s$^{-1}$) is $\sim 40$~Mpc. 
As described in Sect.\,3, the velocity range studied in this work extends up to 
$v \sim 6700$~km\,s$^{-1}$, which corresponds to $\lambda \sim$ 1243 \AA.
Note that distance estimates to galaxies within 3000  ~km\,s$^{-1}$ in the V8k catalogue 
are adjusted to match the Virgo-flow model by \citet{shaya95}.
The relatively uniform sky coverage (except for the zone 
of avoidance, ZOA) of the V8k survey combined with the broad range of galaxy types make it 
suitable for qualitative work \citep{Courtois13}. 

The distribution of apparent and absolute $B$-band magnitudes as well as 
log($L/L^*$) for all galaxies of the V8k catalogue is presented in Fig.~\ref{histV8k}. 
As can be seen from this distribution, the V8k catalogue is largely insensitive 
to dwarf galaxies with luminosities log($L/L^*)\leq -0.5$.
This needs to be kept in mind for our later discussion of the absorber-galaxy relation in Sects. 5 and 6.
We decided to not add supplementary galaxy data from other surveys, because the sky coverage of 
such a mixed galaxy sample would be quite inhomogeneous, which would introduce an additional 
bias to the galaxy-absorber statistics.
 
\begin{figure*}[htp]
\centering
\includegraphics[width=1.\textwidth]{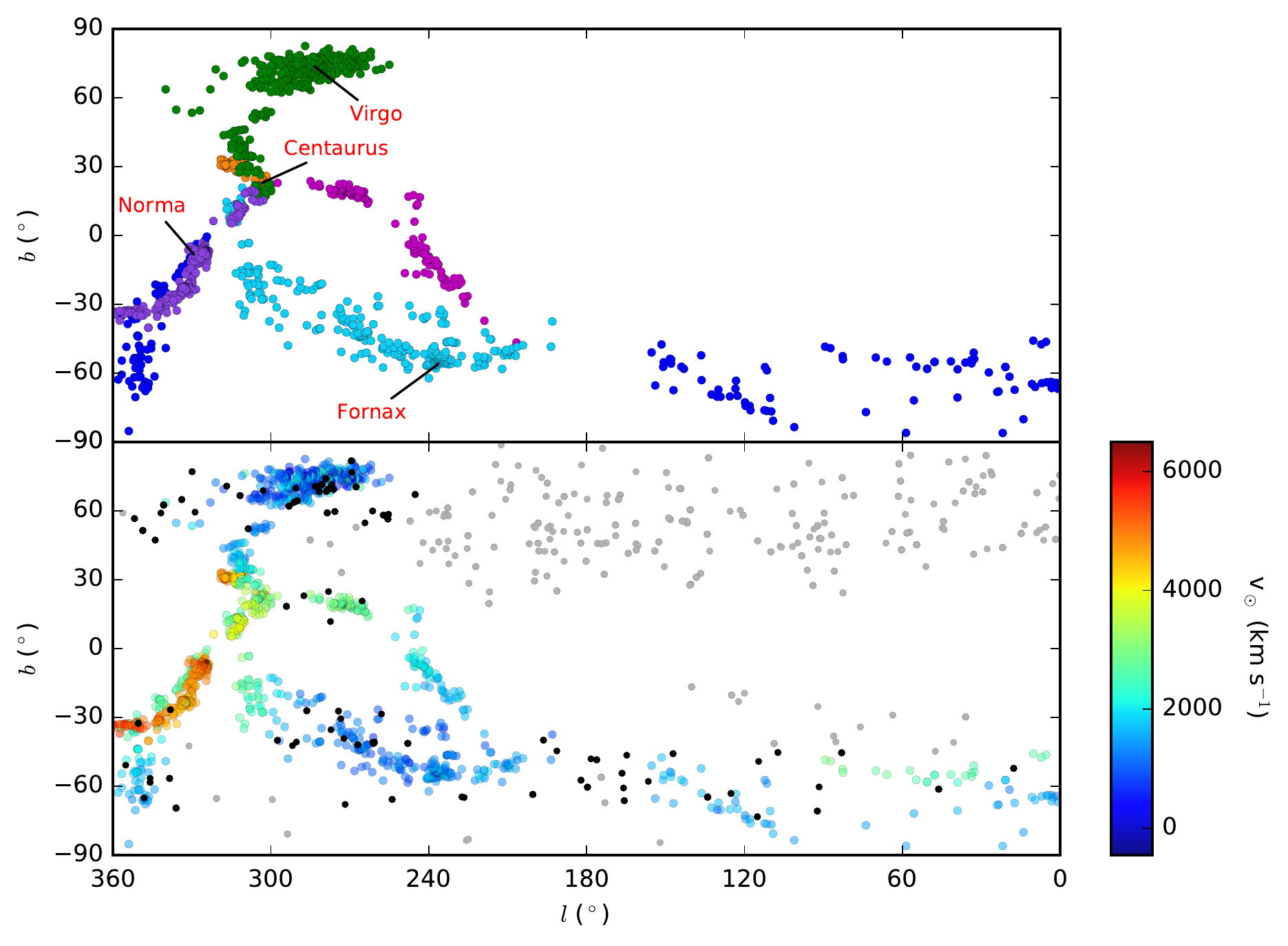}
\caption{
{\it Upper panel:} Sky distribution of galaxies from V8k belonging to filaments as defined 
in \citet{Courtois13}. The different colours indicate different galaxy filaments. Several 
important clusters are noted. {\it Lower panel:} Sky distribution of HST/COS sightlines 
passing close to a filament (black circles) and HST/COS sightlines not belonging to a filament 
(grey circles) plotted together with the galaxies from the V8k catalogue belonging to 
filaments (colour-coded according to velocity).
}
\label{skydistV8k}
\end{figure*}

In Fig.~\ref{skydistV8k}, upper panel, we show the sky distribution of the galaxies in the 
various filaments, such as defined in \citet{Courtois13}.
The galaxies in these filaments have radial velocities in the range
$v=750 - 5900$~km\,s$^{-1}$. All filaments are feeding into the Centaurus Cluster located at 
$l\sim 300\degree$ and $b\sim 20\degree$. The large concentration of galaxies in the green filament,
between $l\sim 260-300\degree$ and $b\sim 60-70\degree$ is due to the Virgo Cluster.

%__________________________________________________________________

%__________________________________________________________________

%%%%%%%%%%%%%%%%%%%%%%%%%%%%%%%%%%%%%%%%%%%%%%%%%%%%%%%%%%%%%%%%%%%%%%%%

\section{Absorption line data}

\subsection{HST/COS observations}

In this study we make use of ancillary HST/COS data, as retrieved from the HST Science Archive 
at the Canadian Astronomy Data Centre (CADC). The total sample consists of 302 AGN sightlines, 
all reduced following the procedures described in \citet{Richter17}. 

Since the Ly$\alpha$ absorption ($\lambda_0=1215.67$ \AA) in the spectra studied here 
falls in the wavelength range between 1215 and 1243 \AA, we make use of the data from the 
COS G130M grating. This grating covers a wavelength range from 1150~$-$~1450~\AA~and has a 
resolving power of $R=16,000-21,000$ \citep{Green2012,Dashtamirova19}. 
The data quality in our COS sample is quite diverse,
with signal-to-noise (S/N) ratios per resolution element
varying substantially (between $3$ and $130$; see Fig.\,A.1 in the Appendix.)

We also checked for metal absorption in the Ly$\alpha$ absorbers, considering the
transitions of Si\,{\sc iii} $\lambda 1206.50$, Si\,{\sc ii} doublet $\lambda 1190.42; 
1193.29$, Si\,{\sc ii} $\lambda 1260.42$, Si\,{\sc ii} $\lambda 1526.71$, Si\,{\sc iv} 
doublet $\lambda 1393.76; 1402.77$, C\,{\sc ii} $\lambda 1334.53$, C\,{\sc iv} doublet 
$\lambda 1548.20; 1550.77$. For the lines at $\lambda > 1450$~\AA, data from the COS 
G160M grating was used, which covers $\lambda$~=~1405~$-$~1775 \AA.

The QSO sightlines are plotted on top of the V8k galaxy filaments in the lower panel of 
Fig.~\ref{skydistV8k}. The sky coverage of the sightlines is noticeably better in the 
upper hemisphere. As can be seen, the majority of the sightlines do not pass through the 
centres of the filaments, but rather are located at the filament edges.

A reason for sightlines not going directly through filaments could be because of extinction. 
This holds true especially for dense regions like the Virgo Cluster, where the extinction is high. 
On the other hand, the Virgo Cluster is a nearby cluster and might be better studied than 
random regions on the sky. From the COS sightlines shown in the lower panel of Fig.~\ref{skydistV8k},
no clear bias can be seen, except for the northern versus southern hemisphere.

\subsection{Absorber sample and spectral analysis}

For all 302 COS spectra, the wavelength range between $1220-1243$ \AA~was inspected for 
intervening absorption. This range corresponds to Ly$\alpha$ in the velocity range 
$v \approx 1070-6700$~km\,s$^{-1}$. At velocities $<1070$~km\,s$^{-1}$, Ly$\alpha$ 
absorption typically is strongly blended with the damped Ly$\alpha$ absorption trough
from the foreground Galactic interstellar medium (ISM). To ensure consistency, we 
do not further consider any absorption feature below 1220 \AA.

Each detected absorption feature at $1220-1243$ \AA\, was checked to be Ly$\alpha$ 
absorption by ruling out Galactic foreground ISM absorption and other, red-shifted 
lines from intervening absorbers at higher redshift. As for the Galactic ISM 
absorption, this wavelength range contains only the N\,{\sc v} doublet (1238, 1242~\AA) 
and the weak Mg\,{\sc ii} doublet (1239, 1240 \AA) as potential contaminants and 
the regions were flagged accordingly. Potential red-shifted contaminating lines 
that were ruled out include: the H\,{\sc i} Lyman-series up to Ly$\delta$, 
Si\,{\sc iii} (1206.50 \AA), and the two O\,{\sc vi} lines at $1037.62$ and 
$1031.93$ \AA. Whenever possible, we also used the line-list of intergalactic 
absorbers from \citet{Danforth16}, which covers a sub-sample of 82 COS spectra.
All in all, we identify 587 intervening Ly$\alpha$ absorbers along the 302 
COS sightlines in the range $\lambda =1220-1243$ \AA.

For the continuum normalisation and the equivalent width measurements of the detected
features (via a direct pixel integration) we used the {\tt span} code \citep{Richter11}
in the ESO-MIDAS software package, which also provides velocities/redshifts for 
the absorbers. To derive column densities of H\,{\sc i} (and the metal ions) for a 
sub-sample of the identified Ly$\alpha$ absorbers we
used the component-modelling method, as described in \citet{Richter13}. In this
method, the various velocity sub-components in an absorber are consistently modelled
in all available ions (H\,{\sc i} and metals) to obtain column densities ($N$) and
Doppler-parameter ($b$-values) for each ion in each component. Throughout the paper, we give 
column densities in units [cm$^{-2}$] and $b$-values
in units [km\,s$^{-1}$].
The modelling code, that is also implemented in ESO-MIDAS, takes into account the wavelength dependent
line-spread function of the COS instrument. Wavelengths and oscillator strengths 
of the analysed ion transitions were taken from the list of \citet{Morton03}. 

The total sample of 302 COS sightlines was separated into two sub-samples, one with 
sightlines passing close to a filament, and the other with sightlines that do not. 
To account for the occasionally seen large projected widths of the filaments (see, e.g., 
part of the dark blue filament in Fig.~\ref{skydistV8k}) and to be able to map 
also the outer parts of the filaments, a separation of 5 Mpc to the nearest galaxy 
belonging to a filament was chosen as dividing distance in this selection process.
One sightline (towards 4C--01.61) was categorised as belonging to a filament -- although 
its nearest galaxy distance is as large as 7.9 Mpc -- because it passes a filament
that is very poorly populated. In total, our selection processes lead to 91 sightlines 
that are categorised as filament-related, while the remaining 211 sightlines 
are categorised as sightlines that are unrelated to the filaments studied here. 
The total redshift pathlength in our COS data set can be estimated as 
$\Delta z = 0.0189\,N$, with $N$ being the number of sightlines. 
This gives $\Delta z = 1.72$ and $3.99$ for the sightline sample belonging to filaments 
and the one unrelated to filaments, respectively. This will be further discussed in 
Sect.\,\ref{section:statistics}.

Within the for us relevant sub-sample of the filament-related sightlines, 12 spectra 
were unsuited for measurements for absorption-line measurements due to various different
data issues, such as an indeterminable continuum, or heavy blending from various lines. 
Of the remaining 79 spectra, 9 had no Ly$\alpha$ absorption features detected in the 
studied wavelength range. This implies a Ly$\alpha$ detection rate of $\sim$90\,\%
(we will later further discuss the number density and cross section of Ly$\alpha$ 
absorbers in this sample). The signal-to-noise ratios for these 79 spectra
vary between 5 and 92 per resolution element. In this sub-sample of 79 filament-related 
sightlines, we identify 215 Ly$\alpha$ absorption systems that are composed of 
227 individual components. For these 215 (227) absorbers (components), we have 
derived H\,{\sc i} column densities and $b$-values via the component-modelling method,
as described above.

In the other sightline sample, that we categorise as unrelated to the galaxy filaments, 
25 spectra were unsuited for measurements for the same reasons as described above.
Of the remaining 186 spectra, only 24 show no Ly$\alpha$ absorption in the range 
considered above, resulting in a 87\,\% detection rate for Ly$\alpha$ in this sample. 

\begin{figure}[tp]
\centering
\includegraphics[width=\hsize]{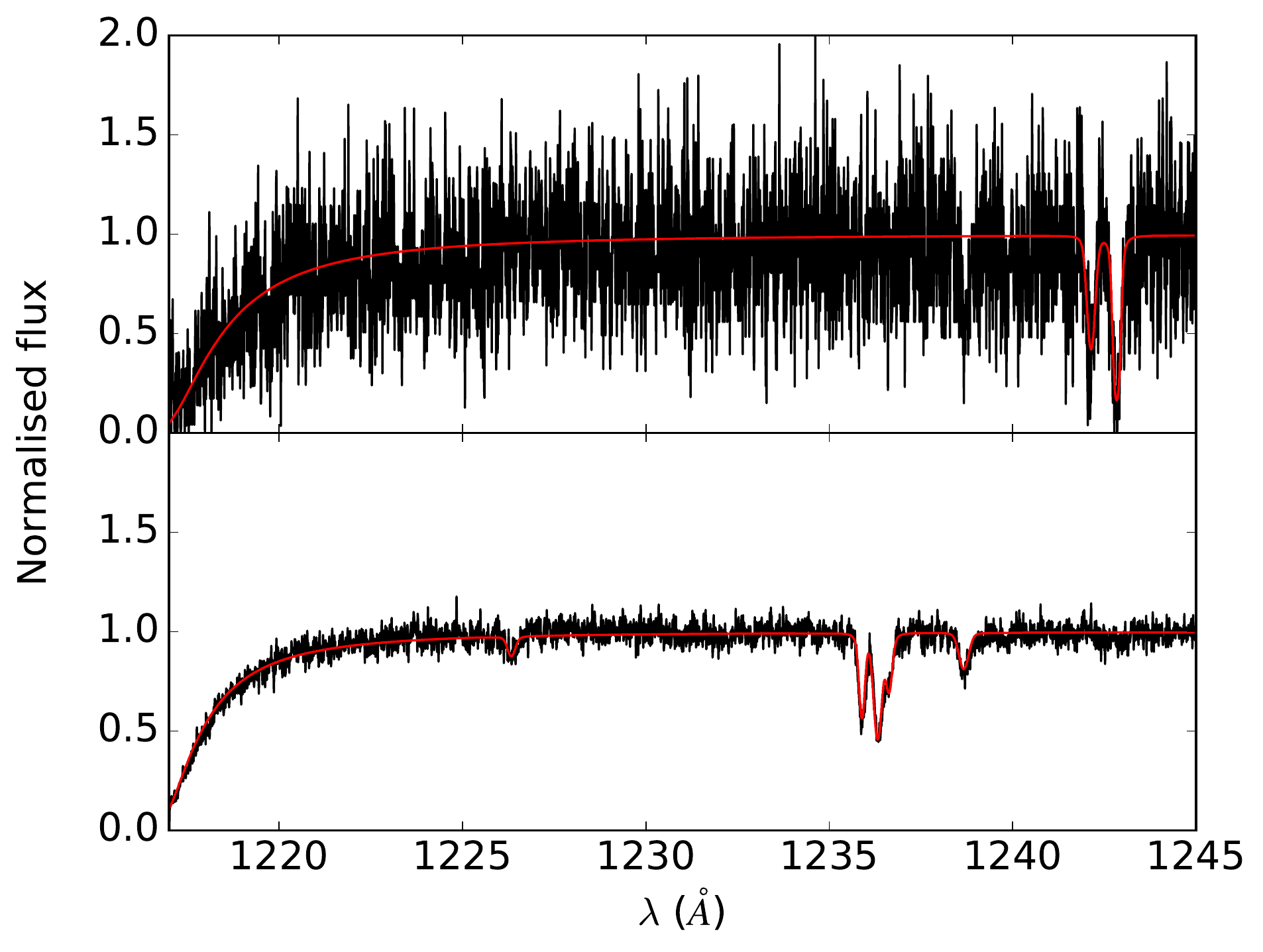}
\caption{
HST/COS G130M spectra of the QSOs VV2006-J131545.2+152556 (upper panel) and 
PKS2155-304 (lower panel). The COS data are given in black, while the absorber 
model is plotted in red. Several Ly$\alpha$ absorbers are seen in these spectra.
For a better visualisation, both spectra are binned over two pixels.}
\label{specmodel}
\end{figure}

Metal ions (Si\,{\sc ii}, Si\,{\sc iii}, Si\,{\sc iv}, C\,{\sc ii} or C\,{\sc iv}) 
were detected for 26 of the 215 Ly$\alpha$ filament absorbers, giving a metal 
detection fraction of $\sim$12\,\%. Two example HST/COS spectra are shown in 
Fig.~\ref{specmodel} (black) together with synthetic model spectrum (red). These
example spectra give an indication of the characteristic differences in S/N in the 
COS data used in this study.

%__________________________________________________________________

%__________________________________________________________________

\begin{figure}[htp]
\centering
\includegraphics[width=\hsize]{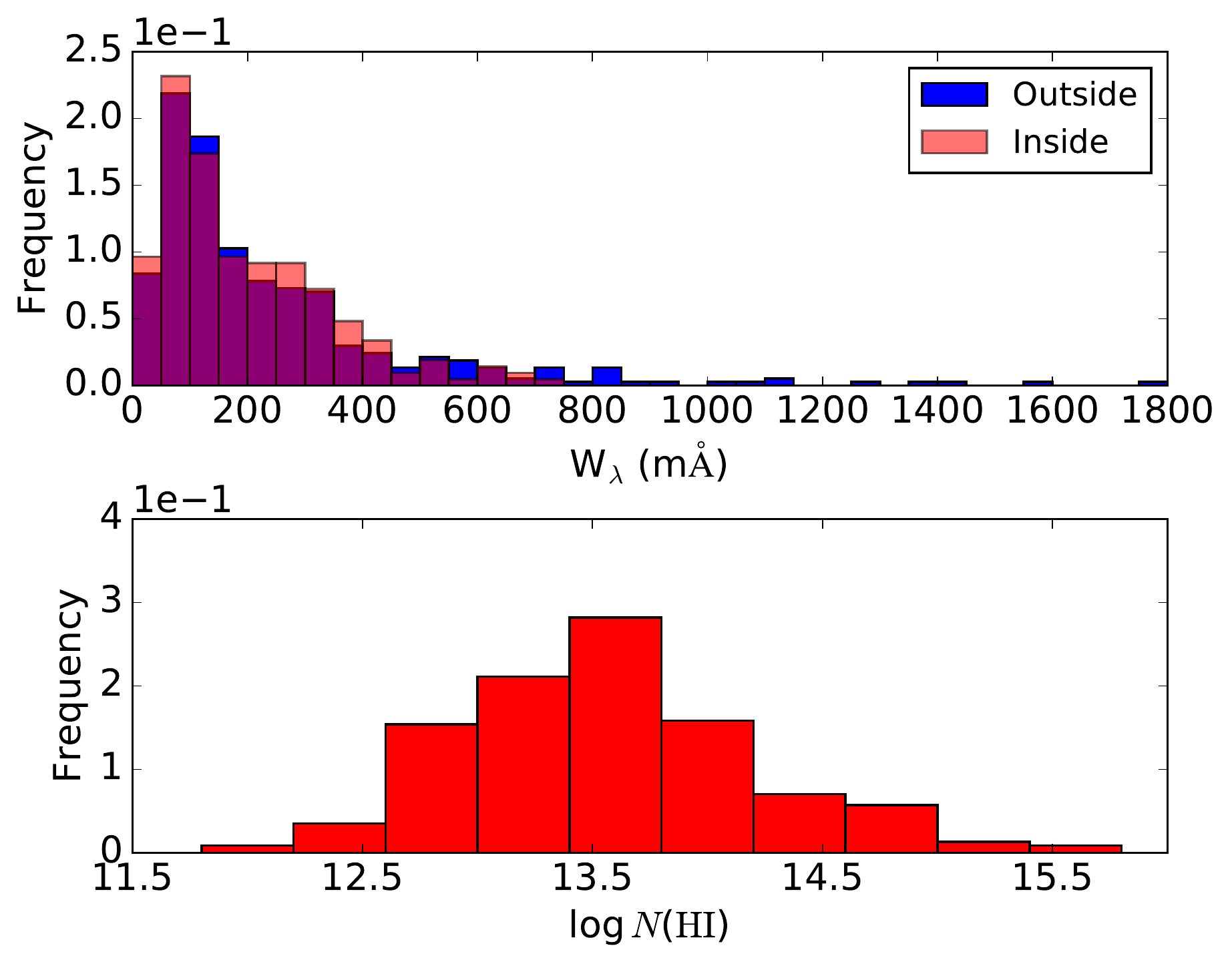}
\caption{
Histogram of equivalent widths of Ly$\alpha$ absorbers (upper panel) and log$N$(H{\sc i}) in 
the filament-related absorbers, as derived from the component modelling (lower panel). 
}
\label{histEWLya}
\end{figure}

Figure~\ref{histEWLya}, upper panel, shows the distribution of H\,{\sc i} Ly$\alpha$ 
equivalent widths for the detected absorbers in the two sub-samples and in the 
combined, total sample. The lower panel instead shows the distribution of H\,{\sc i} 
column densities in the filament-related absorbers, as derived from the component
modelling. Both distributions mimic those seen in previous Ly$\alpha$ studies
at $z=0$ \citep{Lehner07}. The sample of \citet{Danforth16} with 2577 Ly$\alpha$ 
absorbers obtained with HST/COS shows a similar distribution with a peak in equivalent 
width just below 100 m\AA. The H\,{\sc i} column-density distribution falls
off below log $N$(H\,{\sc i}$)=13.5$ due to the incompleteness in the data
to detect weaker H\,{\sc i} Ly$\alpha$ absorbers. Note that because of the
limited spectral resolution and S/N many of the broader Ly$\alpha$
lines most likely are composed of individual, unresolved sub-components.
The H\,{\sc i} column-density distribution function will be discussed in 
Sect.\,7.

%__________________________________________________________________
 
%__________________________________________________________________

Errors of the measured equivalent widths have been derived
with the {\tt span} code \citep{Richter11}, which takes into account 
the S/N around each line, the uncertainty for the local continuum 
placement, and possible blending effects with other lines/features.
Typical 1$\sigma$ errors for the equivalent widths lie around 20 m\AA.
The errors in the column densities were derived based on the component-modelling
method \citet{Richter11}. Here, the typical errors are on the order
of $\sim 0.1$ dex.

The latter value is similar to the errors found by \citet{Richter17} for the same method and a comparable COS data set. The Doppler 
parameters have a relatively high uncertainty,  especially for higher 
values of $b$. With the majority of $b$-values falling between 10 and 
30~km\,s$^{-1}$, the errors are typically $\sim 5$~km\,s$^{-1}$, with 
lower errors for the low end of the range of $b$ and slightly higher 
errors for larger $b$-values. Tabulated results from our 
absorption-line measurements can be made available on request.

%%%%%%%%%%%%%%%%%%%%%%%%%%%%%%%%%%%%%%%%%%%%%%%%%%%%%%%%%%%%%%%%%%%%%%%%

\section{Characterisation of galaxy filaments}

\begin{figure*}[htp]
\centering
\includegraphics[width=0.8\textwidth]{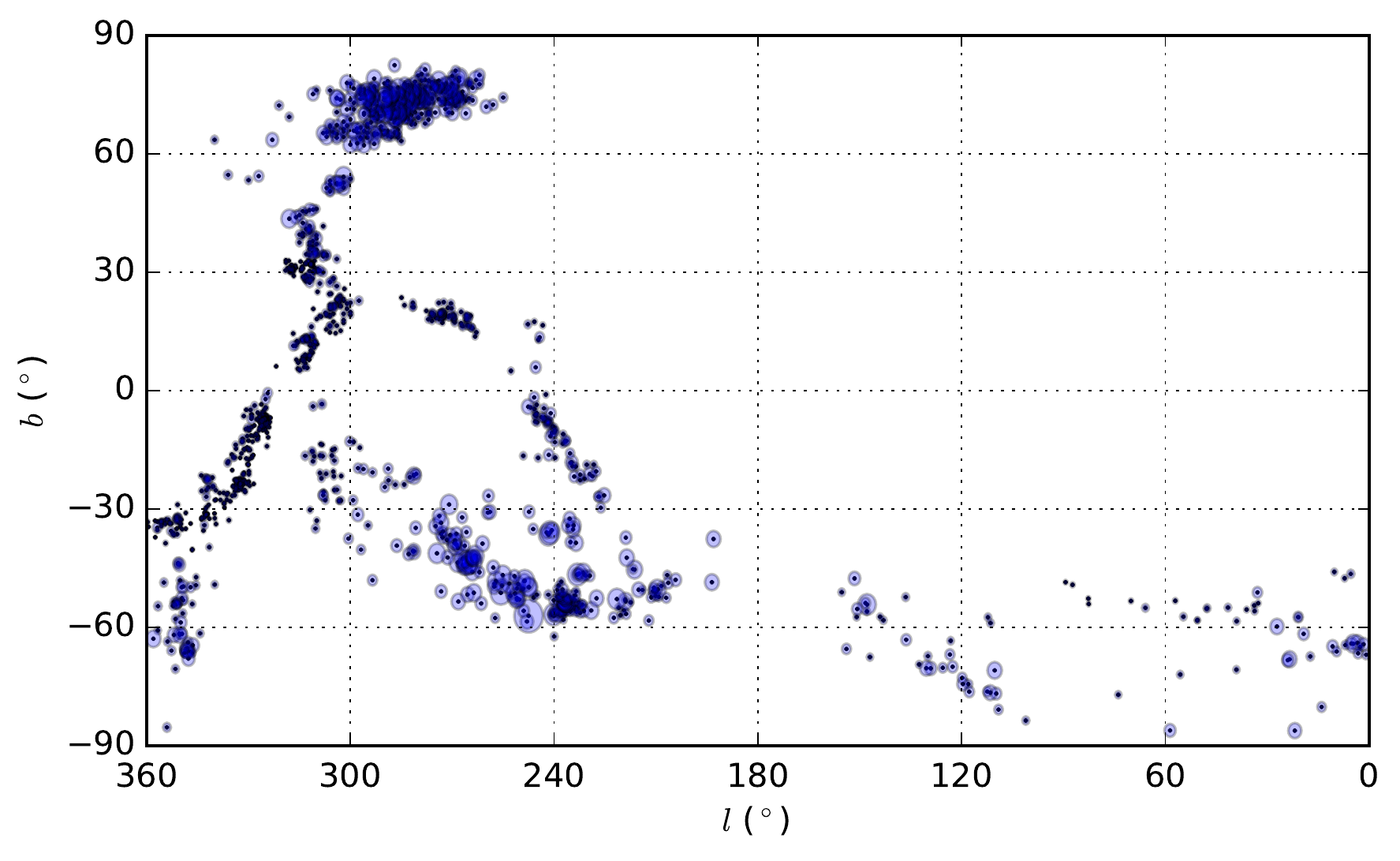}
\caption{
Galaxies belonging to all filaments considered in this study plotted together 
with their projected 1.5 virial radii.
}
\label{skyRvir}
\end{figure*}

To study how the IGM is connected to its cosmological environment, it is 
important to characterise the geometry of the filaments, their galaxy content, 
and their connection to the overall large scale structure. In Fig.~\ref{skyRvir}, 
we show the position of the galaxies in the filaments together with their 
radial extent in 1.5 virial radii (1.5 R$_{\rm vir}$). Gas within this characteristic 
`sphere of influence' can be considered as gravitationally bound to that galaxy.
This plot therefore gives a first indication of how much uncovered 
sky there is {\it between} the galaxies and their spheres of influence, 
indicative for the projected intergalactic space in the filaments (compared 
to the projected circumgalactic space within 1.5 R$_{\rm vir}$). The Virgo Cluster 
clearly stands out, as many galaxies are overlapping in their projected spheres 
at 1.5 R$_{\rm vir}$, while in most other filaments, there are both regions with 
strong overlap and regions without overlapping halos. In Sect.\,6 and in the 
Appendix, we will discuss also other virial radii as selection criteria.

%__________________________________________________________________

%__________________________________________________________________

\begin{figure*}[htp]
\centering
\includegraphics[width=1.0\textwidth]{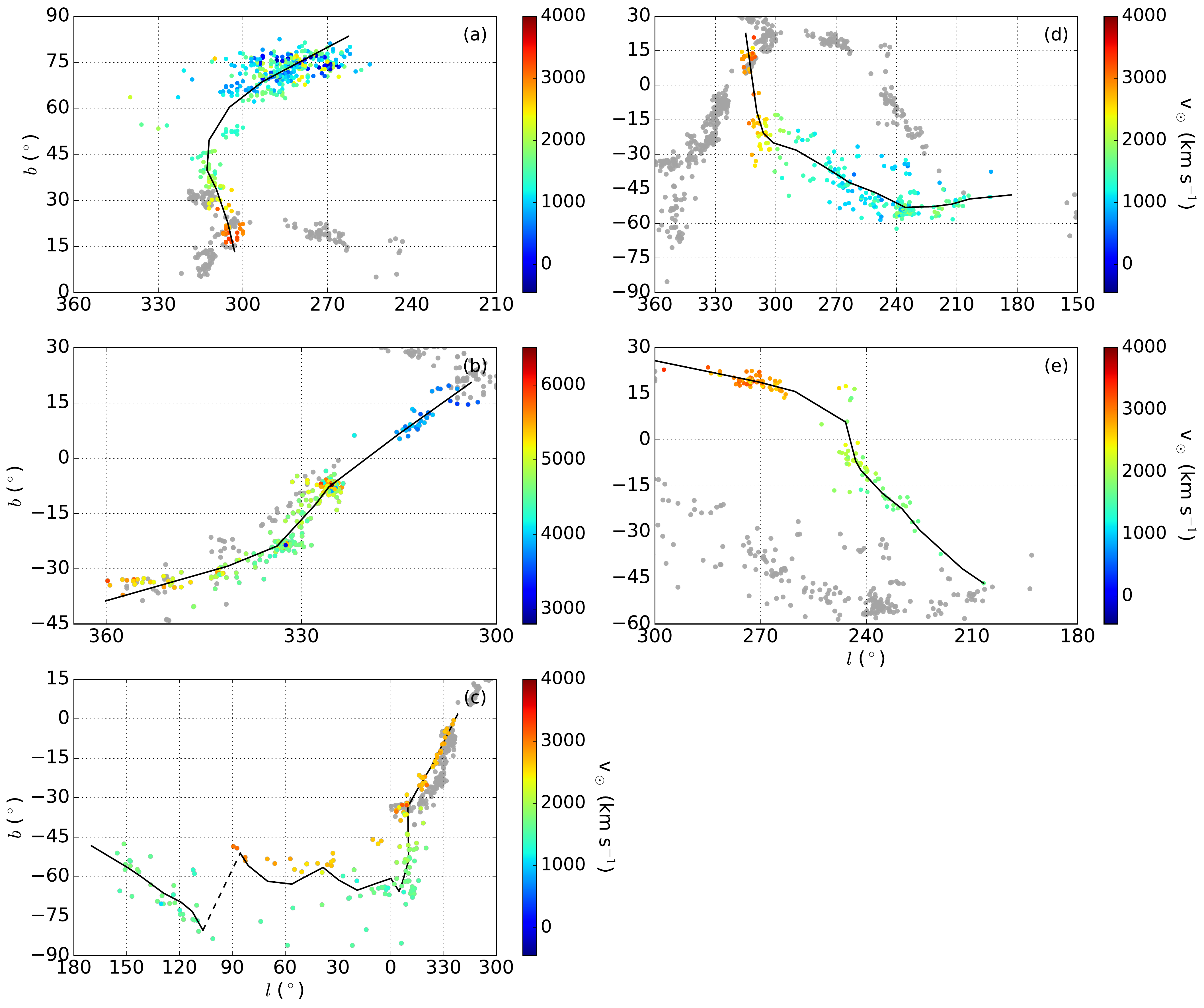}
\caption{Galaxies belonging to different filaments [(a): green; (b): purple; 
(c): dark blue; (d): cyan; (e): magenta] with their velocities 
colour-coded. Grey dots show galaxies belonging to one of the other filaments. 
The filament axes is indicated with the black solid line.
}
\label{filsv}
\end{figure*}

%__________________________________________________________________

\subsection{Parametrisation of filament geometry}

To define an axis for each filament, a rectangular box was generated 
per filament containing the galaxies therein.
The dark blue filament (see Fig.~\ref{skydistV8k}) 
was split into two individual boxes because of geometrical reasons. 
Widths and lengths of the boxes vary for the different filaments, 
as they scale with the filament's projected dimensions.

After defining the boxes sampling the individual filaments, they were 
each sub-divided into segments with the full width of the box and
a length corresponding to 20$\degree$ on the sky. Each segment overlaps 
with the previous one with half the area (10$\degree$ length).
The average longitude and latitude of the galaxies within each segment
was then determined and used as an anchor point to define the filament 
axis. All these anchor points were connected in each filament to form 
its axis.

In this way, the definition of the filament axis on the sky allowed us to calculate
impact parameters of the COS sightlines to the filaments. In addition,
we calculated velocity gradients in the filaments, by taking the average 
velocity of all galaxies in each segment as velocity anchor point.

The method of using overlapping segments to determine the filament axis is 
similar to the approach used by \citet{Wakker15}. A difference with their 
approach is that they first determined which galaxies were part of the 
filament by looking at the velocities. We did not do this as the filaments
were already defined by \citet{Courtois13}. The uncertainty on the placement
of the filament axes is no more than 1.5$\degree$ on the sky, less for most filaments. 

The characteristics of each filament will be discussed separately in the 
following subsections. The orange or `4 clusters' filament from 
\citet{Courtois13} is not discussed here, as there are no available COS 
sightlines nearby. 

\subsection{Green filament}

%__________________________________________________________________

\begin{figure}[htp]
\centering
\includegraphics[width=\hsize]{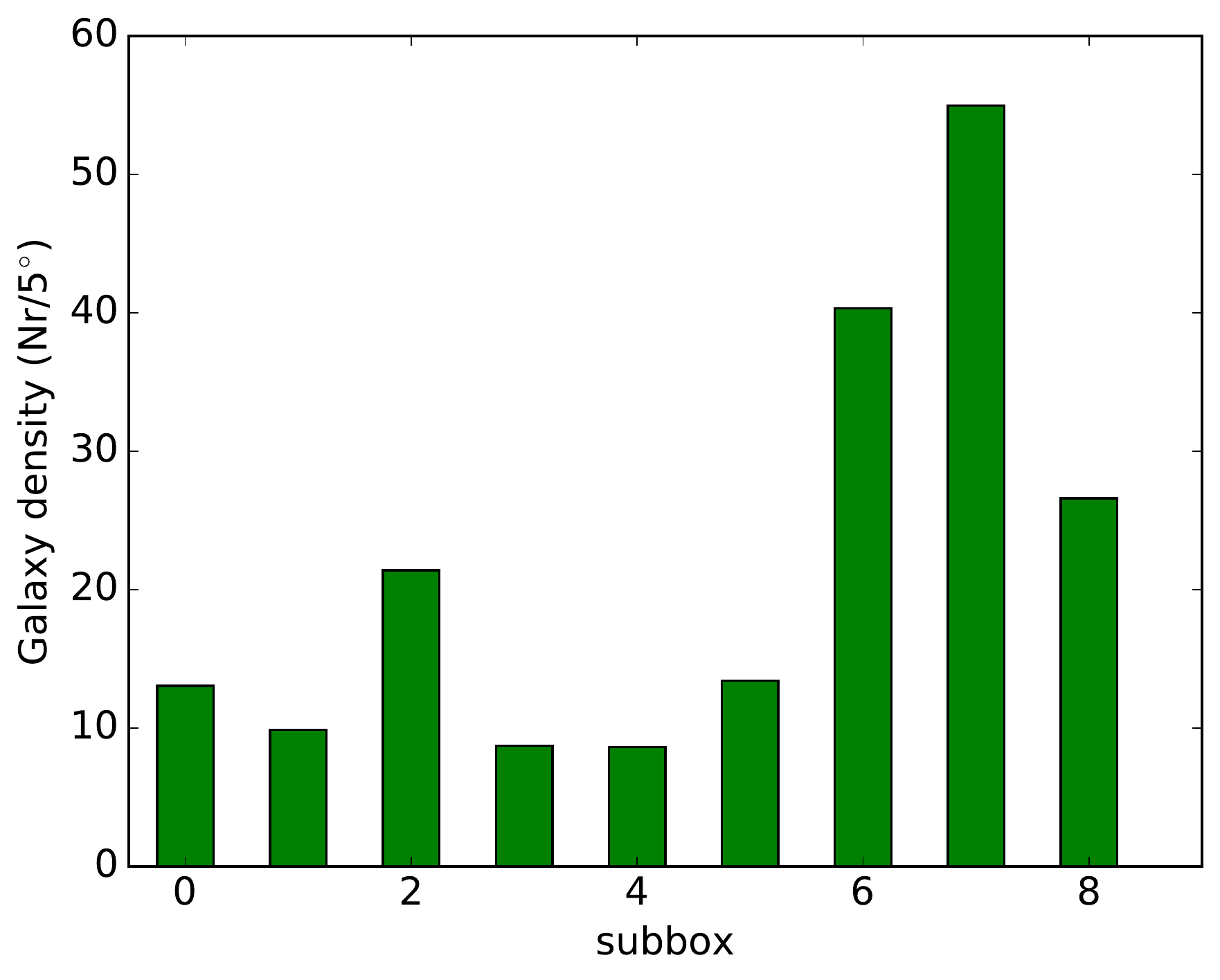}
\caption{Galaxy density along the green filament. The galaxy density indicates
the number of galaxies within 5$\degree$ on the sky for each galaxy.
}
\label{galdensgreen}
\end{figure}

%__________________________________________________________________

Perhaps the most notable of the filaments discussed here is the one 
containing the Virgo Cluster, located at a distance of $\sim$16.5~Mpc \citep{Mei07}
and with up to 2000 member galaxies. 
This filament is labelled in green in 
Fig.~\ref{skydistV8k} and extends from the Centaurus Cluster to the 
Virgo Cluster in the range $l \sim 260-300\degree$ and 
$b\sim 60-70\degree$. The Virgo area has the highest 
galaxy density of the regions studied here. 

The axis of the green filament as well as the galaxy velocities are
indicated in Fig.~\ref{filsv}a. The velocities range from
$\sim 3400$~km\,s$^{-1}$ at the Centaurus Cluster to $\sim-400$~km\,s$^{-1}$
However, the velocities of the galaxies in the Virgo Cluster reach up to 
$\sim 2500$~km\,s$^{-1}$, indicating a large spread in velocities, 
just as expected for a massive galaxy cluster.

Figure~\ref{galdensgreen} shows that the density along the filament varies
greatly, with the Virgo Cluster being the densest region (subboxes 6 $-$ 8).
In total, this filament has 427 galaxies and 36 COS sightlines passing through it.

\subsection{Purple filament}

As mentioned in \citet{Courtois13}, the purple filament is the longest 
cosmological structure in space from those studied here. In projection, 
however, it is one of the shorter filaments on the sky. This filament
was discussed in detail by \citet{Fairall98}, who named it the `Centaurus Wall'. 
A striking lack of galaxies in the regions around $b\sim 0\degree$ is evident
in Fig.~\ref{filsv}b due to the ZOA caused by 
the Milky Way disk in the foreground. Just below this scarcely populated 
region is the Norma Cluster ($l\sim 325\degree$), followed by the Pavus~II 
Cluster ($l\sim 335\degree$).

The purple filament contains the galaxies with the highest velocities in
our sample, with $v$ reaching up to $6500$~km\,s$^{-1}$ (see Fig.~\ref{filsv}b). These high
velocities indicate distances of $\leq 85$ Mpc.
It is the only filament in which the galaxy velocities strongly increase when moving
away from Centaurus. As such, it extends beyond the velocity range considered in 
\citet{Courtois13}. Here, we consider only the part of the filament indicated 
by their work.

The purple filament is the densest of our defined filaments, which is not surprising as
it hosts two galaxy clusters and the projection effect makes it visually compact
on the sky. A total of 351 galaxies from the V8k catalogue belong to this filament,
but only 2 COS sightlines, which are both shared with the dark blue filament.

\subsection{Dark Blue filament}

The dark blue filament represents one branch of the Southern Supercluster filament, 
defined in \citet{Courtois13}. Since it is clearly separated on the sky from the 
other branch (the cyan filament), these two branches are treated as individual 
filaments in this study. Starting from the Centaurus Cluster, the dark blue 
filament is entangled with the purple filament, but it continues to stretch out 
as a rather diffuse cosmological structure over the range $l \sim 0-180\degree$ 
in the southern hemisphere. Because of the low galaxy density, the filament 
axis of the dark blue filament is not well defined and unsteady compared to other 
filaments, as can be seen in Fig.~\ref{filsv}c. The dashed portion of the axis 
indicated in the figure is a result of the small number of galaxies found in 
this region, so the exact filament geometry in this part of the filament 
remains uncertain.

Figure~\ref{filsv} further indicates that average velocities in the dark blue 
filament are much lower than in the purple filament, making the two filaments
easy to distinguish. The dark blue filament also exhibits two 
distinct velocity branches: one with velocities $\sim 2500$~km\,s$^{-1}$ and 
one with v $\sim 1300$~km\,s$^{-1}$ (see Fig.~\ref{filsv}), further 
underlining the inhomogeneous morphology of this filament. This filament has
only 180 galaxies and 21 COS sightlines.

\subsection{Cyan filament}

The second branch of the Southern Supercluster filament is indicated by the 
cyan colour in Fig.~\ref{skydistV8k}. Compared to the dark blue filament, this 
branch is rather densely populated and the corresponding filament axis is
well defined (Fig.~\ref{filsv}d).

As with the green and dark blue filaments, the highest velocities in the cyan
filament are found near the Centaurus Cluster, with velocities decreasing as one
gets closer to the Fornax Cluster. However, Fig.~\ref{filsv} suggests that there 
is a slight increase in velocity near the end of the filament at $l<240\degree$.

The cyan filament is made up of 289 V8k galaxies and there are 20 COS sightlines
passing though it.

\subsection{Magenta filament}

This filament (magenta coloured in Fig.~\ref{skydistV8k}) contains the 
Antlia Cluster and also crosses the ZOA. While it is densely populated 
for $b>0\degree$ (near Centaurus), it is underdense near the ZOA and also only 
moderately populated at negative Galactic latitudes. This makes the transition 
of the filament axis from positive to negative latitudes hard to define. 

As can be seen in Fig.~\ref{filsv}e, the velocities in this filament range 
from 3000 km\,s$^{-1}$ near the Centaurus Cluster to 1400 km\,s$^{-1}$ 
near its end at $l=210\degree$ and $b=-45\degree$. 
It has 143 galaxies and 2 usable COS sightlines.

%__________________________________________________________________

\begin{figure}[tp]
\centering
\includegraphics[width=\hsize]{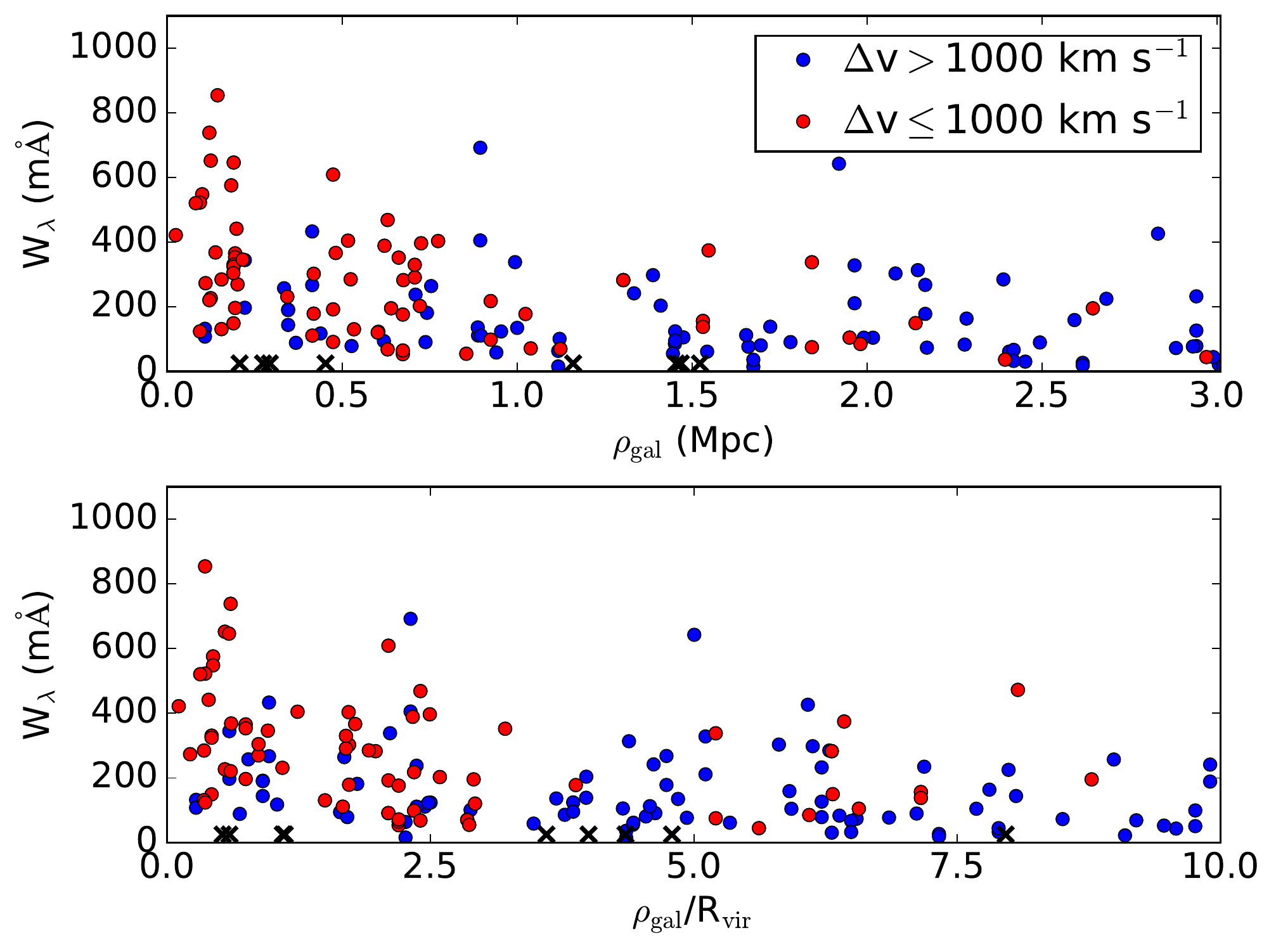}
\caption{
Equivalent width of Ly$\alpha$ absorbers (blue dots) plotted against 
the impact parameter to the nearest galaxy (upper panel) or against 
the impact parameter in units of the galaxy's virial radius (lower panel). 
The sample has been split into components that lie within 1000~km\,s$^{-1}$ 
of the nearest filament segment (red) and  ones with a larger velocity 
difference (blue). Black crosses indicate sightlines that exhibit 
no significant Ly$\alpha$ absorption 
in the analysed spectral region.
For these, we give the distance to the nearest galaxy in the velocity 
range $v~=~1070-6700$~km\,s$^{-1}$.
}
\label{EWrhog}
\end{figure}

%__________________________________________________________________

%%%%%%%%%%%%%%%%%%%%%%%%%%%%%%%%%%%%%%%%%%%%%%%%%%%%%%%%%%%%%%%%%%%%%%%%

\section{Ly$\alpha$ absorption and its connection to galaxies}\label{Lyagals} 

To learn about the relation between intervening Ly$\alpha$ absorption, 
nearby galaxies, and the local large scale structure, in which the 
absorbers and galaxies are embedded, we first look at the connection 
between Ly$\alpha$ absorption in the COS data and individual galaxies.

In Fig.~\ref{EWrhog}, upper panel, we have plotted the equivalent widths 
of all Ly$\alpha$ absorbers against the line-of-sight impact parameter to 
the nearest galaxy, $\rho_{\rm gal}$, that has a radial velocity within 
400~km\,s$^{-1}$ of the absorber. For this plot, {\it all} V8k galaxies 
have been taken into account (not just the ones in filaments), as some of 
the absorbers might be related to galaxies outside of the main cosmological
structures. We indicate absorbers that are within 1000~km\,s$^{-1}$ of
the nearest filament in red, and those that have larger deviation
velocities in blue.
Non-detections have been indicated by the black
crosses. The corresponding sightlines do not show any Ly$\alpha$ absorption 
in the wavelength range 1220$-$1243 \AA.

There is an overdensity of absorbers within 1 Mpc of the nearest 
galaxy, many of which having equivalent widths less than $W${\boldmath$_{\lambda}$} $<200$ m\AA. 
This overabundance of weak absorbers close to galaxies might be a selection 
effect. Prominent regions, such as the dense Virgo Cluster, receive more 
attention by researchers and are sampled by more sightlines (and by spectral
data with better S/N) compared to underdense cosmological regions, which
typically are not as well-mapped. The highest equivalent widths 
of the absorbers ($W_{\lambda} >500$ m\AA) typically are found closer to the galaxies,
in line with the often observed anti-correlation between Ly$\alpha$ 
equivalent width and impact parameter \citep[e.g.][]{Chen01,French17}. 
There is, however, a large scatter in this distribution, such as seen also 
in other studies \citep[e.g.][]{French17}. This scatter most likely is related 
to filament regions that have a large galaxy density and overlapping (projected) 
galaxy halos, such as indicated in Fig.~\ref{skyRvir}. Ly$\alpha$ absorption
that is detected along a line of sight passing through such a crowded region 
cannot unambiguously be related to a {\it particular} galaxy (such as the 
nearest galaxy, which is assumed here), but could be associated with the same 
likelihood to any other (e.g., more distant) galaxy and its extended
gaseous halo that is sampled by the sightline. 

The lower panel of Fig.~\ref{EWrhog} shows the Ly$\alpha$ equivalent width 
plotted against $\rho_{\rm gal}$/R$_{\rm vir}$. Again, we see the same trend for 
stronger absorbers to be closer to a galaxy. Out of the 208 Ly$\alpha$ 
absorption components, 29 are within $1.5$ virial radii from the nearest 
galaxy. Following \citet{Shull14,Wakker15}, this is the characteristic radius 
up to which the gas surrounding a galaxy is immediately associated with that 
galaxy and its circumgalactic gas-circulation processes (infall, outflows,
mergers). It corresponds to $\sim 2 - 3$ times the gravitational radius as
defined in \citet{Shull14}.
Outside of this characteristic radius, the gas is more likely 
associated with the superordinate cosmological environment (i.e., the group
or cluster environment and the large-scale filament; but see also Sect.\,6 and 
Fig.\,B.1).
\citet{Wakker15} use both this distance criterion and the criterion of 
absorption occurring within 400~km\,s$^{-1}$ of the galaxy's velocity 
to associate each absorber with either the galaxy or the filament.
This velocity range (which we also adopt here; see above) is justified in view
of other dynamic processes that would cause a Doppler shift of the gas
in relation to the galaxy's mean radial velocity, such as galaxy rotation, 
velocity dispersion of gas-structures within the virialised dark-matter of 
the host galaxy, as well as in- and outflows.

\begin{figure}[tp]
\centering
\includegraphics[width=\hsize]{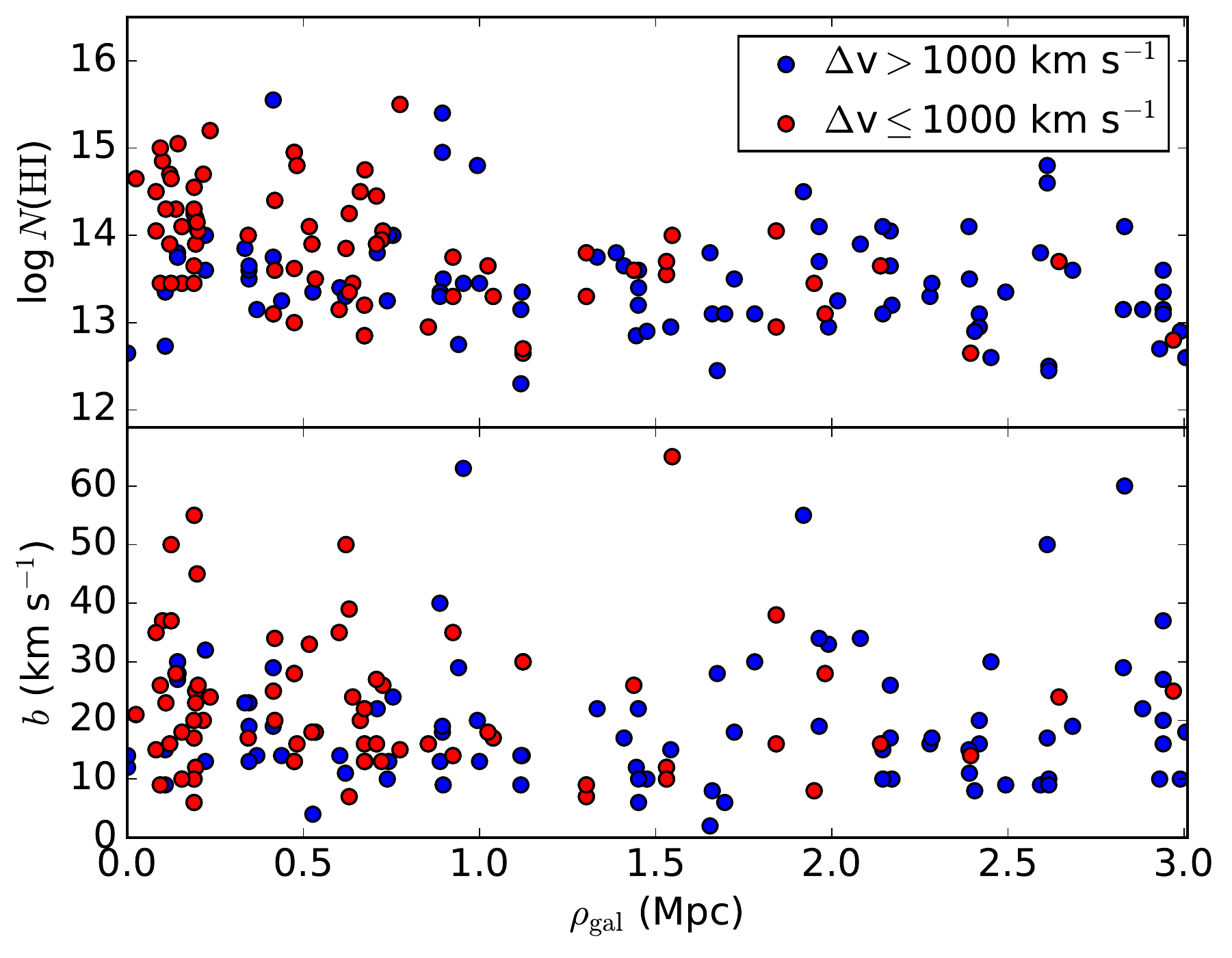}
\caption{
Logarithmic H\,{\sc i} column density and the Doppler parameter of
Ly$\alpha$ absorbers are plotted against $\rho_{\rm gal}$ for the
same two samples as shown in Fig.~\ref{EWrhog}.
}
\label{NHIb}
\end{figure}

In Fig.~\ref{NHIb} we show how the H\,{\sc i} column density 
(log$N$(H\,{\sc i})) and the Doppler parameter ($b$ value) vary with
$\rho_{\rm gal}$. Similarly to $W${\boldmath$_{\lambda}$}, the largest 
values for log$N$(H\,{\sc i}) and $b$ are found at smaller impact parameters, 
but (again) the scatter is large.

\citet{Wakker15} have also plotted the equivalent width versus impact parameter
to the nearest galaxy for their sample. Although there are some high equivalent
width absorbers at large $\rho_{\rm gal}$ (out to 2000 kpc), 
the average equivalent width decreases with
increasing $\rho_{\rm gal}$. Similar to our sample, \citet{Wakker15} find the 
majority of the absorbers within 1 Mpc of a galaxy. Our sample, however, has
a larger scatter and more strong absorbers at larger distances. 
\citet{Prochaska11} also conclude there is an anti-correlation between
equivalent width and galaxy impact parameter for their sample that
has a maximum $\rho_{\rm gal}$ of 1 Mpc. In addition to stronger absorbers
having lower impact parameters, their sample shows an increase of
the number of weak absorbers ($W_{\lambda} <$ 100 m\AA) 
with increasing impact parameter.

%__________________________________________________________________

%__________________________________________________________________

\begin{figure*}[htp]
\centering
\includegraphics[width=1.0\textwidth]{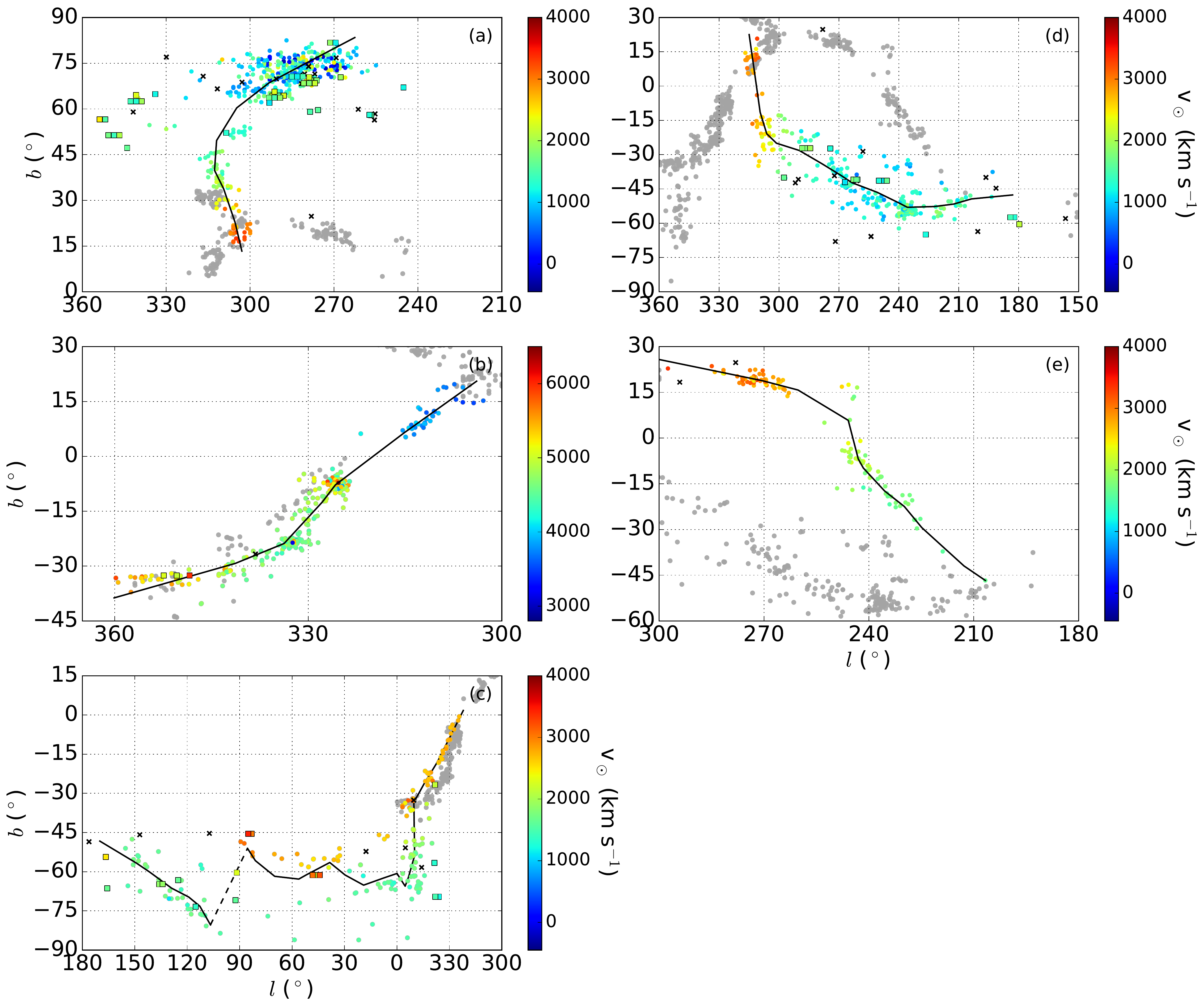}
\caption{
Same as Fig.~\ref{filsv}, but here with Ly$\alpha$ absorbers (coloured 
squares) overlaid that fall within 1000~km\,s$^{-1}$ of the filament's 
velocity. Multiple absorbers along the same sightline have been given 
spatial offset. COS sightlines that do not exhibit Ly$\alpha$ absorption 
in this range, are indicated by black crosses.
}
\label{filsCOS}
\end{figure*}

%__________________________________________________________________

%%%%%%%%%%%%%%%%%%%%%%%%%%%%%%%%%%%%%%%%%%%%%%%%%%%%%%%%%%%%%%%%%%%%%%%%

\section{Ly$\alpha$ and its connection to filaments}

In Fig.~\ref{filsCOS}, the filaments are plotted together with the position 
of the COS sightlines (filled squares) and the velocities of the detected 
Ly$\alpha$ components colour-coded (in the same way as the galaxies).
Only those absorption components are considered that have velocities within 
1000~km\,s$^{-1}$ of the nearest filament segment. These plots are useful 
to visualise the large-scale kinematic trends of the absorption features 
along each filament, while at the same time the spatial and kinematic 
connection between Ly$\alpha$ components and individual galaxies can
be explored.

In the green filament (a), Ly$\alpha$ absorption is predominantly found 
near 1500~km\,s$^{-1}$. This holds true for both the sightlines at the outskirts
of the filament and those going through the Virgo Cluster. For the latter, this
indicates the gas has a higher velocity than the typical velocity
of galaxies in the Virgo Cluster (as mentioned earlier, the V8k catalogue 
takes into account the Virgo-flow model by \citet{shaya95}).
Due to the extended Ly$\alpha$ trough of the Galactic foreground ISM 
absorption, intervening Ly$\alpha$ absorption below $\sim 1100$~km\,s$^{-1}$ 
cannot be measured in our COS data set, so that our absorption statistics is
incomplete at the low end of the velocity distribution. Still, 
the trend of decreasing galaxy velocities with increasing distance to the 
Centaurus Cluster (see above) is not reflected in the kinematics of the 
detected Ly$\alpha$ absorbers in this filament, which appears 
to be independent of the large-scale galaxy kinematics.

The purple filament (b) and first section of the dark blue filament (c) 
overlap on the sky and have two COS sightlines in common. 
The different filament velocities allow us to assign the detected Ly$\alpha$ 
absorption in one of the sightlines to the purple filament, while the other 
sightline has one absorption component that we associate with the dark blue 
filament. With only 2 sightlines available for the purple filament, no 
clear trends can be identified.

As the dark blue filament continues, the different `branches' noted earlier 
in Sect.\,4.4 are also reflected in the velocities of the Ly$\alpha$ 
absorption components. This trend might be partly a result of our original
selecting criterion for filament-related absorbers (absorption within 1000 km\,s$^{-1}$
of the closest filament-segment velocity; see above). However, because of the 
large velocity range used, the selection criterion cannot account for the 
entire branching effect. Obviously, in this filament, the gas traces the 
velocities of the galaxies. Since this is the most diffuse filament, the 
chance of finding a Ly$\alpha$ absorber, that is not directly associated 
with a galaxy but rather traces the large-scale flow of matter in that 
filament, is higher.

The cyan filament (d) instead is well-populated with galaxies, while also 
being relatively long and broad. It thus has a high cross-section and there
are several sightlines that pass through this structure. Also in this case,
the Ly$\alpha$ absorption appears to follow the velocity trend of the 
galaxies in the filament here. Starting from Centaurus, the absorbers
first exhibit velocities around 1800~km\,s$^{-1}$, then the velocities
decrease several hundred km\,s$^{-1}$, to rise again slightly at the 
end of the filament, in line with the galaxies' velocity pattern.

Most of the sightlines that pass the magenta filament (e) are not suited 
for a spectral analysis. The one sightline that has been analysed, shows 
no significant absorption in the relevant velocity range, implying that
no useful information is available for the magenta filament.

%__________________________________________________________________

\begin{figure}[tp]
\centering
\includegraphics[width=\hsize]{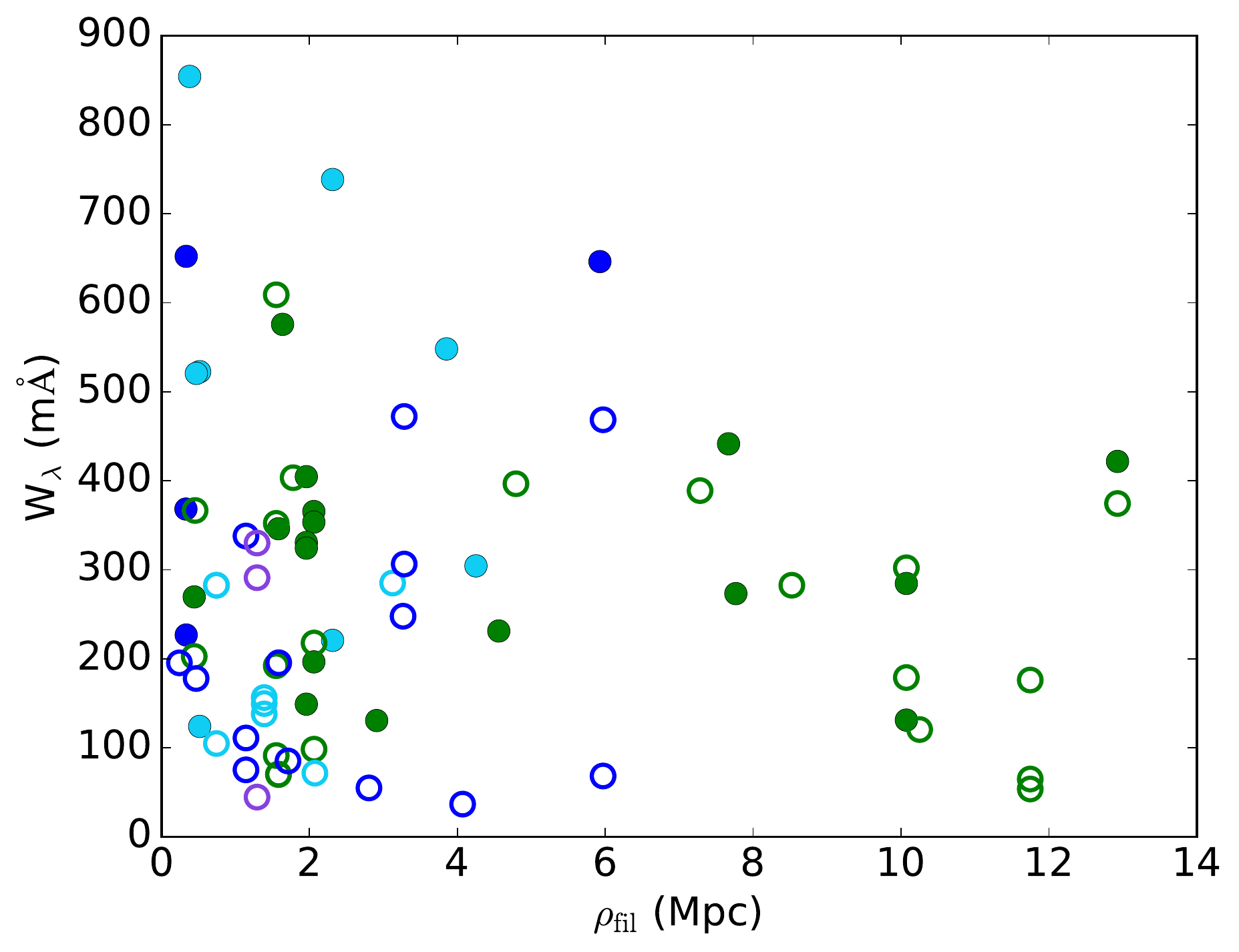}
\caption{
Ly$\alpha$ equivalent width versus filament impact parameter for absorbers 
with velocities within 1000~km\,s$^{-1}$ of the nearest filament segment. 
Open circles indicate absorbers that are associated with a galaxy (passing 
it within 1.5 R$_{\rm vir}$ and $\Delta$v < 400~km\,s$^{-1}$), closed circles are 
not associated with a (known) galaxy. The different colours indicate the 
individual filaments.
}
\label{EWrhof}
\end{figure}

%__________________________________________________________________

In analogy to Fig.~\ref{EWrhog}, Fig.~\ref{EWrhof} shows the equivalent 
width of the Ly$\alpha$ absorbers, but now plotted against the {\it filament} 
impact parameter, $\rho_{\rm fil}$. To evaluate whether the absorbers are related 
to a nearby galaxy, those absorbers that pass within 1.5 R$_{\rm vir}$ and 
$\Delta$v~<~400~km\,s$^{-1}$ of a galaxy are shown as open circles, whereas 
absorbers not associated with a (known) galaxy are indicated with closed 
circles. In the Appendix we show in Fig.\,\ref{diffrhog} the effect of 
varying the impact-parameter criterion for absorbers to be associated with a galaxy 
between $1.0$ and $2.5$ R$_{\rm vir}$ (which leads to no further insights, however).

While some of the absorbers with the highest equivalent widths are associated 
with a galaxy, this is not true for all strong absorbers. Neither sub-sample 
shows a clear, systematic trend for the equivalent width scaling with 
$\rho_{\rm fil}$, except that the maximum Ly$\alpha$ equivalent width 
in a given $\rho_{\rm fil}$ bin decreases with in increasing distance. However, 
both sub-samples show a higher absorber density within $\rho_{\rm fil}<3$~Mpc 
compared to more distant regions. Some of the absorbers indicated in green
extend up to $\rho_{\rm fil}=13$~Mpc, but these absorbers are unlikely to be 
part of the green filament, as the typical width of a cosmological filament 
is a few Mpc \citep{Bond10}. 
But even if we limit our analysis to absorbers with $\rho_{\rm fil} < 5$~Mpc 
\citep[as in][]{Wakker15}, the large scatter in the distribution
of Ly$\alpha$ equivalent widths versus filament impact parameter remains.

\begin{figure*}[ht!]
\centering
\includegraphics[width=0.8\textwidth]{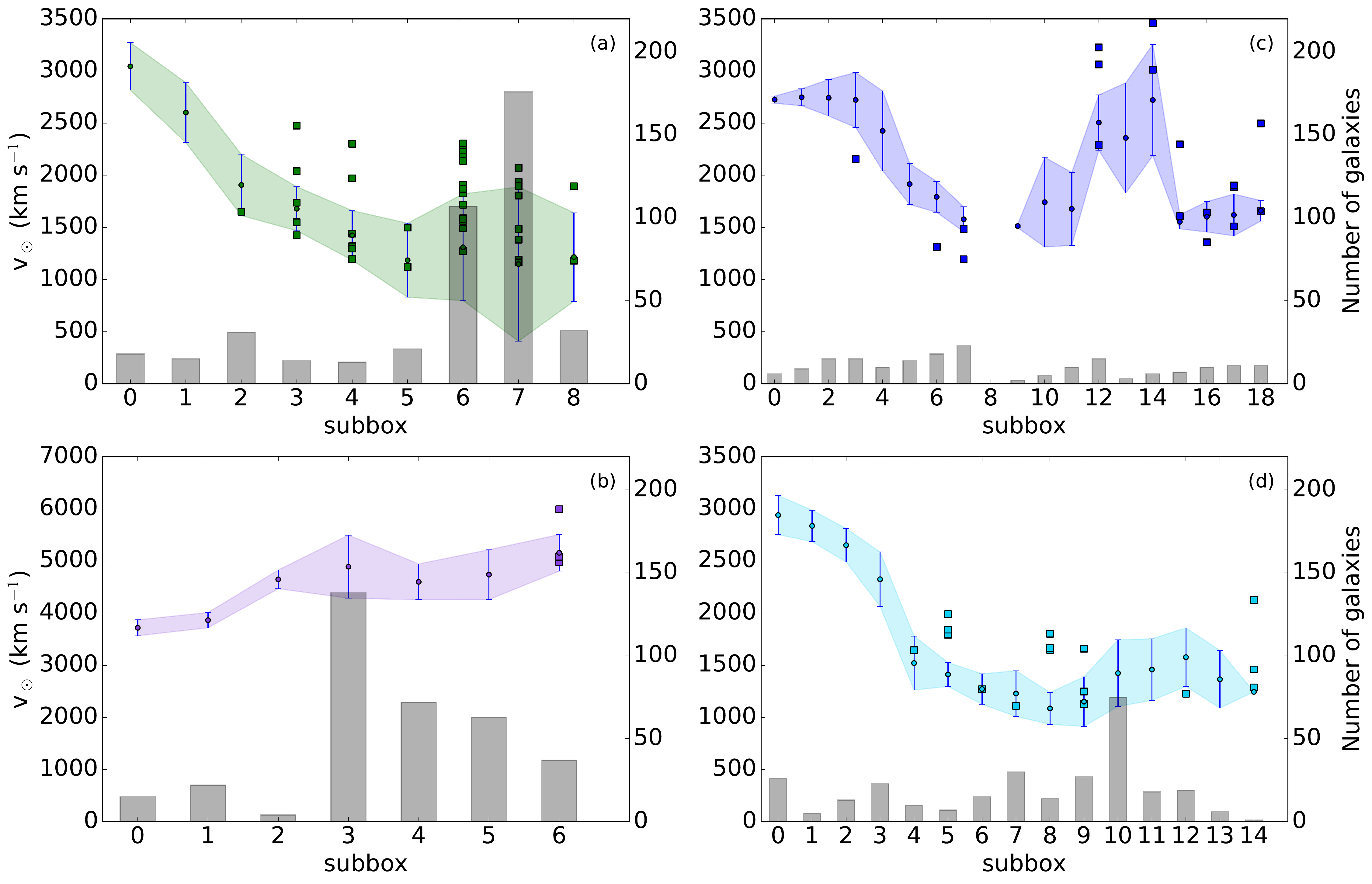}
\caption{
Average galaxies velocities along four filaments (dots, (a): green; 
(b): purple; (c): dark blue; (d): cyan) plotted together with the velocities of 
Ly$\alpha$ absorbers (squares) for each sub-box (segment). The velocity 
dispersion is indicated by the colour-shaded area. The grey bars indicate 
the numbers of galaxies belonging to each sub-box (segment) in the filament.
}
\label{kinbins}
\end{figure*}

The velocity trends for galaxies and absorbers along four filaments (green, 
purple, dark blue, cyan)  are shown in Fig.~\ref{kinbins}. 
Starting point for each filament is the Centaurus-Cluster region. 
Here, each sub-box (segment) is defined to have  a length of 10$\degree$ 
on the sky. This is half the length of the sub-boxes (segments) used to 
define the filament axes (see Sect.\,4.1), because here, sub-boxes (segments)
do not overlap. Only for the second part of the dark blue filament 
(sub-boxes 12$-$18), a length of 20$\degree$ was chosen to have a 
sufficient number of galaxies available for the determination of a 
meaningful average velocity.

%__________________________________________________________________

%__________________________________________________________________

\begin{figure}[tp]
\centering
\includegraphics[width=\hsize]{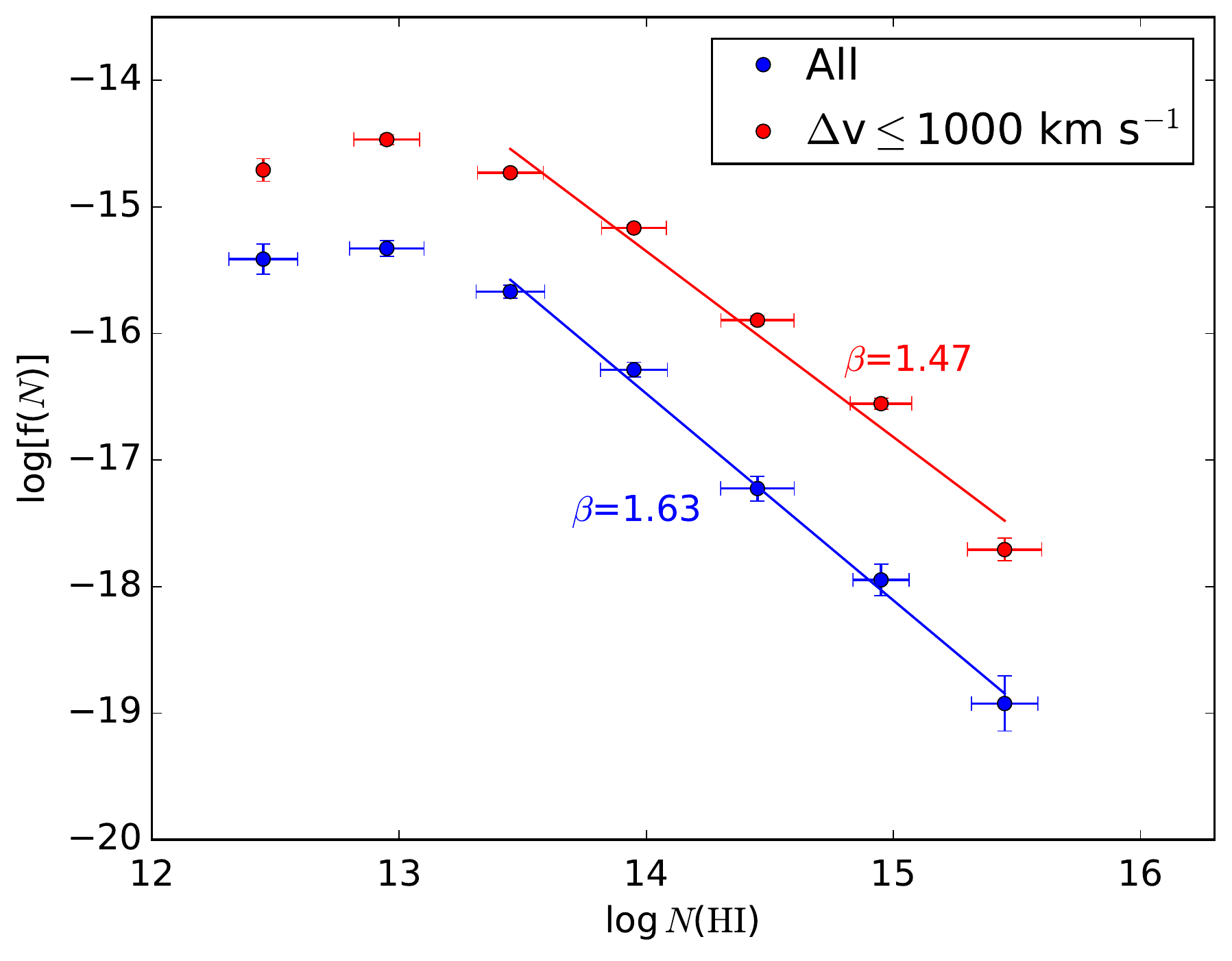}
\caption{
H\,{\sc i} column-density distribution function for the Ly$\alpha$ 
absorbers in sightlines close to filaments (blue) and the absorbers 
falling within 1000~km\,s$^{-1}$ from the filament velocity (red). 
Errors in log[f($N$)] are from Poisson statistics.
}
\label{CDDF}
\end{figure}

%__________________________________________________________________

\begin{figure}[tp]
\centering
\includegraphics[width=\hsize]{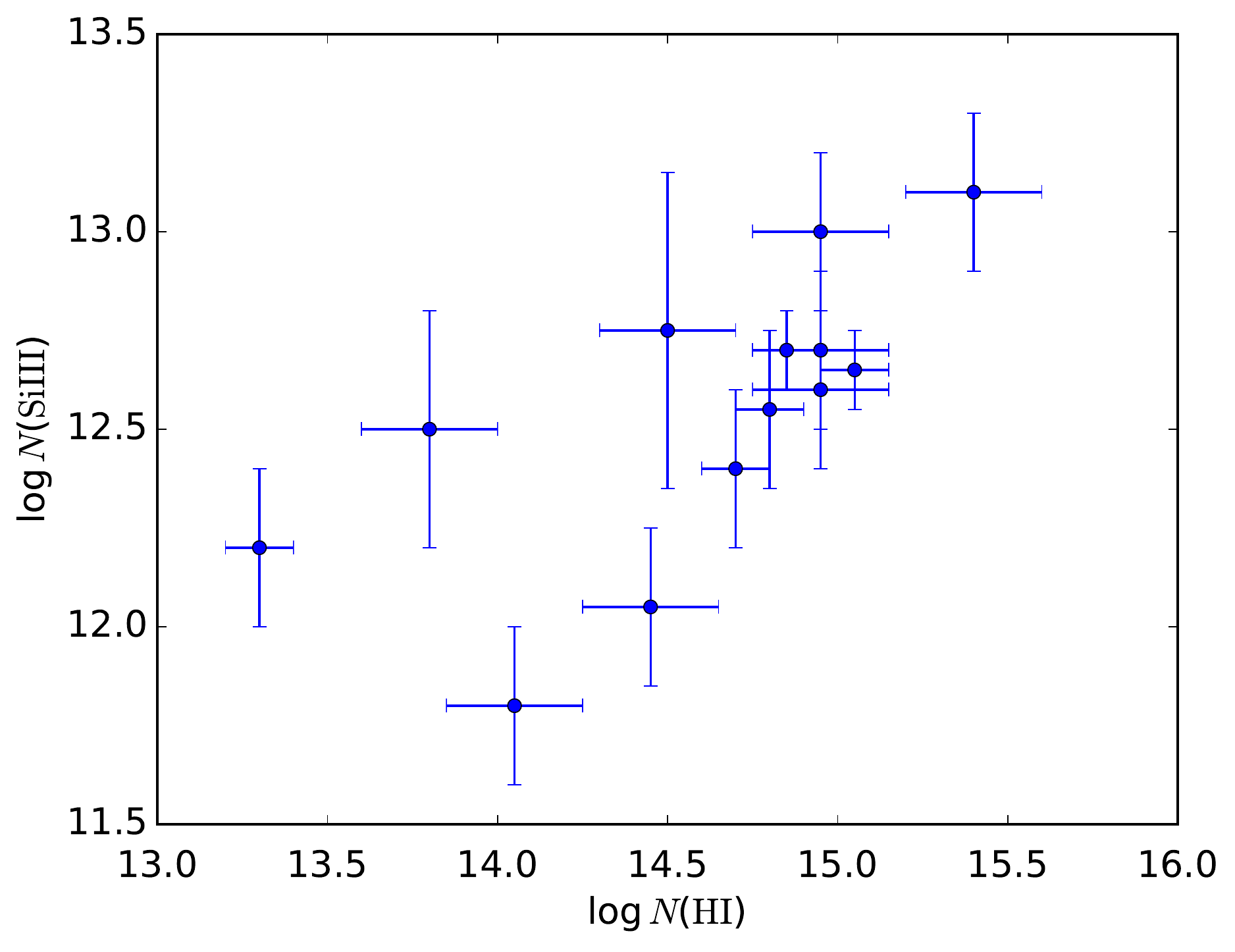}
\caption{
Relation between log$N$(Si\,{\sc iii}) and log$N$(H\,{\sc i}) for 
13 systems in our absorber sample.}
\label{SiIIIHI}
\end{figure}

%__________________________________________________________________

Both the green and cyan filaments (Fig.\ref{kinbins}a and d) show a clear 
decrease in velocity as they extend further away from the Centaurus Cluster. 
For these two filaments, the velocities of the detected Ly$\alpha$ absorbers 
all lie above the lower limit of the galaxy velocity-dispersion in each sub-box,
with only one exception (see sub-box 3 in the green filament, where one absorber 
falls just below the shaded area). This could possibly mean that there is a void
of absorbers in the region between the filament and the Milky Way. 
However, it is important to recall that absorbers with velocities less than 
$\sim$1100~km\,s$^{-1}$ could not be measured due to the Galactic foreground 
absorption. This limit could also explain why there is no absorption in the 
lower velocity range of the Virgo Cluster (see sub-boxes 6 and 7 in 
Fig.~\ref{kinbins}a).

Furthermore, most absorbers in sub-boxes 2 $-$ 4 and a couple in the other 
sub-boxes have large $\rho_{\rm fil}$ ($>5$~Mpc). The velocity spread for those
absorbers in the first sub-boxes is larger and trending to slightly higher
velocities than that of the galaxies. Most, however, fall within the 
standard deviation of the galaxy-velocities.

The purple filament contains only three absorbers, all belonging to the same 
sightline at the end of the filament, as can be seen in Fig.~\ref{kinbins}b. 
The plot underlines that this filament is well-populated, with more than 
50 galaxies in 4 out of 7 sub-boxes. 

Just like the filament axis in Fig.~\ref{filsv}c, the velocity trend along 
the dark blue filament is irregular. This trend is also reflected in the 
absorber velocities. While there are only a few galaxies in each sub-box, 
the second part of the filament contains 14 absorbers, which is comparable 
to the total number of absorbers in the cyan filament.

%%%%%%%%%%%%%%%%%%%%%%%%%%%%%%%%%%%%%%%%%%%%%%%%%%%%%%%%%%%%%%%%%%%%%%%%

\section{Absorber statistics}
\label{section:statistics}

In quasar-absorption spectroscopy, the observed relation between the number 
of H\,{\sc i} absorption systems in the column density interval $\Delta N$ 
($N_{\mathrm{H\,I}}$ to $N_{\mathrm{H\,I}} +dN$) and the absorption-path 
length interval $\Delta x$ ($X$ to $X + dX$) is commonly characterised by 
the differential {\it column density distribution function} (CDDF), 
$f(N_{\mathrm{H\,I}})$. We use the formalism described in \citet[][and 
references therein]{Lehner07} and adopt the following expression to
describe the differential CDDF of our Ly$\alpha$ absorbers:

\begin{equation}
f(N_{\mathrm{H\,I}})\,dN_{\mathrm{H\,I}}\,dX = C_{\mathrm{H\,I}}\,
N_{\mathrm{H\,I}}^{-\beta}\,dN_{\mathrm{H\,I}}\,dX.
\end{equation}

Following e.g. \citet{Tytler87}, absorption path $\Delta x$ and 
redshift path $\Delta z$ at $z\approx 0$ can be approximated by the 
relation 

\begin{equation*}
\Delta X = 0.5[(1+\Delta z)^2 - 1],
\end{equation*}

where we calculate the redshift pathlength $\Delta z$ for the various sightline 
samples as described in Sect.\,3.2.
The slope of the CDDF is given by the exponent $\beta$, while the normalisation 
constant, $C_{\mathrm{H\,I}}$ can be calculated via the relation

\begin{equation*}
C_{\mathrm{H\,I}} \equiv m_{tot}(1-\beta)/\{N_{max}^{1-\beta} 
[1-(N_{min}/N_{max})^{1-\beta}]\}
\end{equation*}

Here, $m_{tot}$ is the total number of absorbers in the column-density 
interval $N_{min}$ to $N_{max}$.

The column density distributions for our sample of Ly$\alpha$ absorbers
are shown in Fig.~\ref{CDDF}. The CDDFs were fitted for Ly$\alpha$ components 
with log $N$(H\,{\sc i}) $\geq 13.2$ (maximum log\,$N$(H\,{\sc i}) = 15.5).
For the full filament sightline sample, we derive $\beta = 1.63 \pm 0.12$, 
while for the sub-sample of absorbers within 1000~km\,s$^{-1}$ of the filament 
velocity we obtain $\beta = 1.47 \pm 0.24$. 
Using high-resolution STIS data, \citet{Lehner07} derived 
$\beta = 1.76 \pm 0.06$ for their sample of narrow 
absorbers (b $\leq$ 40~km\,s$^{-1}$, 110 absorbers) and $\beta = 1.84 \pm 0.06$ 
for b $\leq$ 150~km\,s$^{-1}$ (140 absorbers), thus slightly steeper slopes
than the distributions found here. Note that we do not split our sample based on 
$b$-values, as the fraction of absorbers with $b > 40$~km\,s$^{-1}$ 
is small ($\leq 5\,\%$).

A comparison with other recent studies can be made, for example with the \citet{Danforth16} 
COS study of the low-redshift IGM, which yields $\beta = 1.65 \pm 0.02$ for a 
redshift-limited sub-sample of 2256 absorbers. Other results are: 
$\beta = 1.73 \pm 0.04$ from \citet{Danforth08} (650 absorbers from COS), 
$\beta = 1.65 \pm 0.07$ by \citet{Penton00} (187 absorbers from GHRS 
and STIS), and $\beta = 1.68 \pm 0.03$ by \citet{Tilton12} (746 absorbers 
from STIS). Most of these values are consistent with $\beta = 1.60 - 1.70$, 
whereas higher values ($\beta > 1.70$) may indicate a redshift 
evolution of the slope between $z = 0.0 - 0.4$. Such an evolution was 
discussed in \citet{Danforth16}, who find a steepening of the slope with 
decreasing $z$ in this redshift-range.
For our study, with $z \leq 0.0223$, the slope should be close to the
value valid for the universe at $z = 0$.

Furthermore, the spectral resolution may play a role in the determination of $\beta$.
For instance, the spectral resolution of COS ($R\approx 20,000$) is 
substantially lower than the resolution
of the STIS spectrograph ($R\approx 45,000$) used by \citet{Lehner07},
so that some of our Ly$\alpha$ absorption components might be composed
of several (unresolved) sub-components with lower column densities. 
The limited S/N in the COS data additionally hampers the detection of 
weak Ly$\alpha$ satellite components in the wings of stronger 
absorbers \citep[see also][their Fig.\,1]{Richter06}.

In this series of results, the shallower CDDF ($\beta = 1.47 \pm 0.24$) 
for the sub-sample of velocity-selected absorbers within 1000~km\,s$^{-1}$ 
of the filaments stands out. Although this low value is formally in 
agreement within its error range with the canonical value of $\beta = 1.65$,
it may hint at a larger relative fraction of high-column density systems
in the filaments, reflecting the spatial concentration of galaxies and 
the general matter overdensity in these structures. A larger sample
of Ly$\alpha$ absorbers associated with filaments would be required to 
further investigate this interesting aspect on a statistically 
secure basis.

Like previous studies, our sample offers an opportunity to study 
the number of Ly$\alpha$ absorbers per unit redshift ($d\mathcal{N}/dz$). 
Table~\ref{table:linedens} gives the Ly$\alpha$ line density for 
the entire sample, as well as for several subsamples. Those subsamples 
separate the sample into different column-density bins, allowing us to 
directly compare the results to the high-resolution \citet{Lehner07} absorber 
sample and other studies.

For the full absorber sample (including filament and non-filament related 
sightlines), the Ly$\alpha$ line density is $116\pm5$ lines 
per unit redshift, but only for the filament-related subsample do we have
column-density information available (see Sect.\,3.2).
Taking this subsample in the ranges $13.2\lesssim$ 
log$N$(H\,{\sc i}) $\lesssim 14.0$ and $13.2\lesssim$ 
log$N$(H\,{\sc i}) $\lesssim 16.5$, we derive Ly$\alpha$ line densities
of $88\pm 8$ and $98\pm 8$, respectively. These values are in
good agreement with those reported by \citet{Lehner07}, who derive
number densities of $80\pm6$ and $96\pm7$ for the same 
column-density ranges. 

Our results can also be compared with those obtained from the much larger 
COS absorber sample presented in \citet{Danforth16}. They derive a 
relation for $d\mathcal{N}/dz$ in the form 
$d\mathcal{N} (>N)/dz \approx 25\, (N/10^{14}$~cm$^{-2})^{-0.65}$. 
For absorbers with log$N$(H\,{\sc i})$\geq 13.2$, this leads to 
$d\mathcal{N}/dz \sim 83$, mildly lower than the values derived 
by us and \citet{Lehner07}, but still in fair agreement.

If we take the velocity-selected absorber sample, which potentially traces 
the Ly$\alpha$ gas associated with the filaments, we obtain a significantly 
higher line density of $d\mathcal{N}/dz=189\pm25$. 
The redshift pathlength for the velocity-selected absorber sample was 
calculated for each sightline by considering a velocity range of 
$\pm 1000$ km\,s$^{-1}$ around the center-velocity 
for the filament segment that was closest to that sightline. 

The value of $189\pm25$ for the velocity-selected filament sample is $93$ percent
higher than the value derived for the total filament-absorber sample (along the 
same lines of sight). 
This line overdensity of the Ly$\alpha$ forest kinematically associated with filaments 
obviously reflects the matter overdensity of baryons in the potential wells 
of these large-scale cosmological structures.

%__________________________________________________________________

\begin{table}
\caption{Ly$\alpha$ line density for the full sample (filament and 
non-filament related sightlines), filament related sightlines, and for the velocity 
selected absorber sample ($\Delta$v < 1000~km\,s$^{-1}$). }      
\label{table:linedens}    
\centering                                  
\begin{tabular}{l r r }     
\hline\hline
 log$N$(H\,{\sc i}) & $\mathcal{N}$ &  $d\mathcal{N}/dz$\\   
\hline
Full sample  & & \\
 & 579 & $116\pm 5$ \\ \hline
Filament sample & & \\
$12.0 - 16.5$ &  215   &  $144\pm 11$ \\
$13.2 - 14.0$ &   132  &  $88\pm 8$   \\
$13.2 - 16.5$ &  147   &   $98\pm 8$  \\ \hline
Velocity-selected filament sample &  &\\
$12.0 - 16.5$ & 74     & $233\pm 27$ \\
$13.2 - 14.0$ & 46     & $145\pm 22$ \\
$13.2 - 16.5$ & 60     & $189\pm 25$ \\
\hline                                         
\end{tabular}
\end{table}

%__________________________________________________________________

For the sake of completeness, we also show in Fig.~\ref{SiIIIHI} the
relation between log$N$(Si\,{\sc iii}) and log$N$(H\,{\sc i}) for absorbers
in our sample for which both species are detected. Only for a small 
fraction (8.4\,\%) of the Ly$\alpha$ components, Si\,{\sc iii} can be 
measured, which is partly because of the velocity-shifted Si\,{\sc iii} 
falling in the range of Galactic Ly$\alpha$ absorption. Generally,
log$N$(Si\,{\sc iii}) increases with log$N$(H\,{\sc i}), as expected
from other Si\,{\sc iii} surveys in the IGM and CGM 
\citep[e.g.][]{Richter16}, but the scatter is substantial. The small 
number of Si\,{\sc iii}/H\,{\sc i} absorbers in our sample does not 
allows us to draw any meaningful conclusions on the metal content 
of the absorbers in relation to their large-scale environment.

%%%%%%%%%%%%%%%%%%%%%%%%%%%%%%%%%%%%%%%%%%%%%%%%%%%%%%%%%%%%%%%%%%%%%%%%

\section{Discussion on Ly$\alpha$ absorbers and their environment}

In their study, \citet{Prochaska11} correlated galaxies and Ly$\alpha$ 
absorbers {\rm at $z=0.06-0.57$} and found that for weak absorbers (13 
$\leq$ log$N$(H\,{\sc i}) $\leq$ 14) less than 20\,\% of the systems were 
associated with a known galaxy, while for strong absorbers 
(log$N$(H\,{\sc i}) $\geq$ 15), this fraction was 80\,\%. The criteria 
they used for associating a galaxy with an absorber were the following: 
i) a velocity difference between absorber and galaxy of 
$\leq$ 400~km\,s$^{-1}$, and ii) an impact parameter of $\rho 
\leq$ 300 kpc. Using the same criteria, we derive for our sample that 
10\,\% (40\,\%) of the low (high) column density absorbers are associated 
with a galaxy in the V8k galaxy sample.

Therefore, and in agreement with \citet{Prochaska11}, we find that 
high column density Ly$\alpha$ absorbers are four-times more often 
associated with a galaxy than low column density absorbers, but the
overall fraction of absorbers for which an associated galaxy is found
is only half of that in the \citet{Prochaska11} sample. This can be 
attributed to the fact that the V8k catalogue is incomplete
for $M_B<-16$ and $v>1000$ km\,s$^{-1}$, while the 
\citet{Prochaska11} galaxy sample is complete up to at least
$z = 0.1$ for galaxies with $L < 0.1 L_{*}$.
By comparing their observed covering fractions with a filament model, 
\citet{Prochaska11} conclude that {\it all} Ly$\alpha$ absorbers are 
associated with either a galaxy or a filament. This view is debated 
by \citet{Tejos12}, however, who argue that there is an additional 
population of `random' Ly$\alpha$ absorbers that reside in the 
underdense large-scale structure (voids).

The idea of Ly$\alpha$ absorbers belonging to different populations 
(and thus different environments) was proposed more than 25 years ago 
by \citet{Morris93}. By analysing Ly$\alpha$ absorbers in a single 
sightline and comparing the location of the absorbers with locations
of galaxies, these authors found that the absorbers do not cluster 
around galaxies as strongly as galaxies cluster among themselves. 
On the other hand, they also found the trend that the absorbers do 
cluster around galaxies. From this, they concluded that there could 
be two populations of Ly$\alpha$ absorbers: one that is associated 
with galaxies and one that is more or less randomly distributed.

To test whether the Ly$\alpha$ absorbers in our sample resemble a 
`random population', we generated two artificial populations of 
Ly$\alpha$ absorbers, both with random sky positions, random 
absorption velocities within the assumed $v_{\rm fil} \pm 1000$~km\,s$^{-1}$ 
velocity range of a filament, and random H\,{\sc i} column densities 
weighted by the H\,{\sc i} CDDF.
For the one population, we have restricted the sample 
from \citet{Lehner07} (hereafter abbreviated with L07) 
to the redshift range spanned by the filaments in our study
and used the slope of their CDDF (resulting in 39 absorbers, 
$\beta=1.76$). For the other population, we used our own absorber sample  
and slope (74 absorbers, $\beta=1.47$).
The normalisation constant and absorber-path length were calculated 
using the relations given above. All absorbers are assumed to be 
at a distance of $\leq 5$ Mpc from the nearest galaxy belonging to 
a filament, which was also our original criterion to select absorbers 
inside a filament. The fraction of the simulated absorbers in each 
filament was adjusted to match the real fractions found in this study. 
Because the dark blue and purple filaments have an overlap on the sky, 
their randomised simulated counterparts were generated for the both 
filaments combined. 

\begin{figure}[tp]
\centering
\includegraphics[width=\hsize]{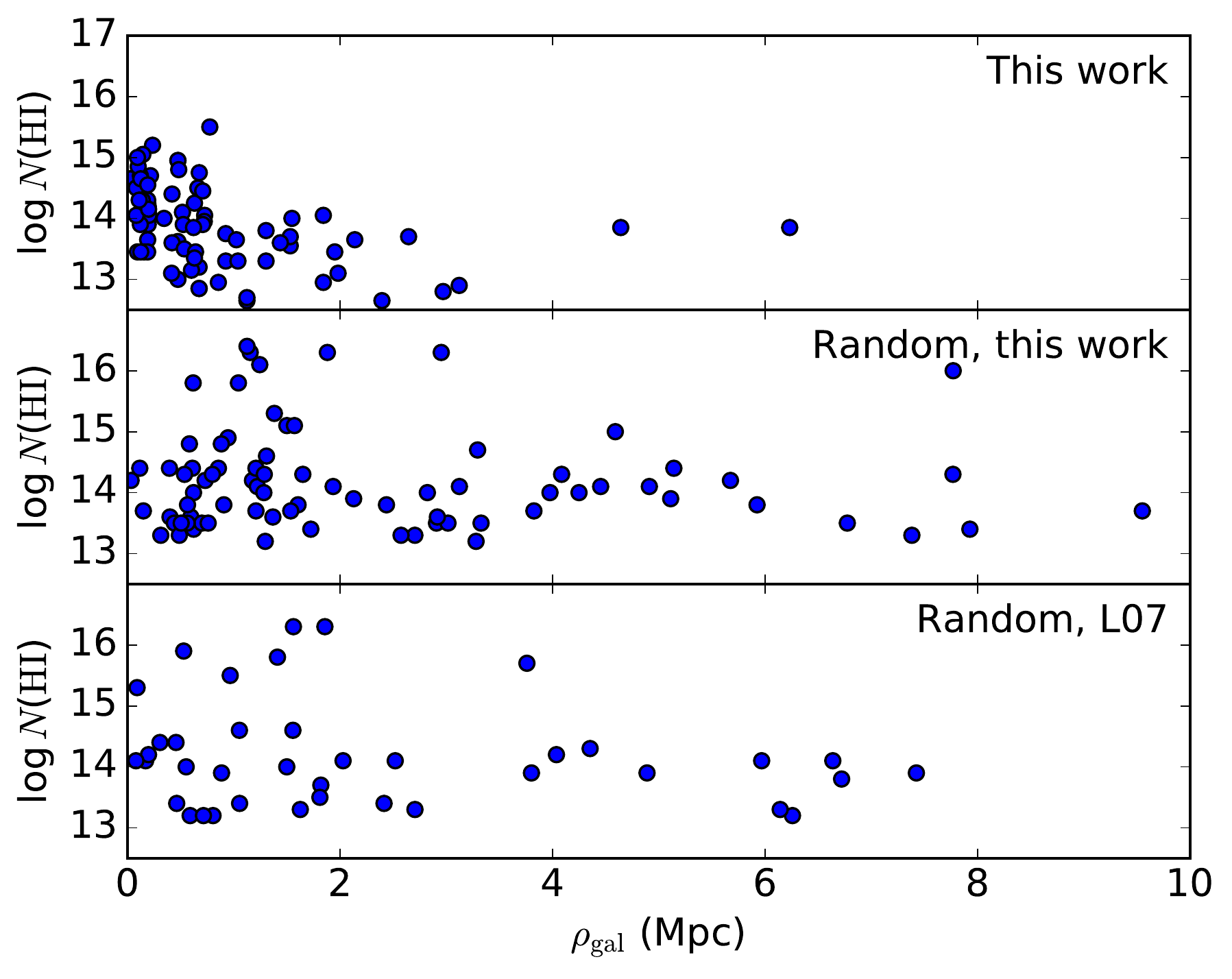}
\caption{
H\,{\sc i} column density versus $\rho_{\rm gal}$ for Ly$\alpha$ absorbers, 
i) as measured in the COS data (upper panel), ii) for a randomised sample 
following the CDDF in this work (middle panel), and iii) for a randomised 
sample following the number statistics and CDDF from \citet{Lehner07}
(lower panel).
}
\label{randomrhog}
\end{figure}

Figure~\ref{randomrhog} shows a comparison of how column densities
for the three different Ly$\alpha$ absorber samples (observed sample, 
random sample with own statistics, random sample with L07 statistics)
depend on $\rho_{\rm gal}$. Like in Sect.~\ref{Lyagals}, $\rho_{\rm gal}$ was 
calculated for the nearest galaxy on the sky within a velocity interval 
of 400~km\,s$^{-1}$ from the absorber. Clearly, the measured absorbers 
cluster more strongly around galaxies than both random samples.
This indicates that at least some of the absorbers are associated with 
galaxies, as expected from previous studies 
\citep[e.g.][]{Morris06,Prochaska11,Tejos14,French17}. 

A very rough estimate of the fraction of absorbers associated with a 
galaxy can be made by comparing the fraction of absorbers within 1.5 Mpc
of a galaxy in different samples. For the measured absorbers 82\,\% of
the absorbers have $\rho_{\rm gal} \leq 1.5$~Mpc, while the fraction for
the randomised sample drawn form our own distribution is 53\,\% and
for the L07 random sample it is 46\,\%. In conclusion, about a third
of our absorbers cannot be explained by a random population
and might be connected to a nearby galaxy.

%__________________________________________________________________

%__________________________________________________________________

\begin{figure}[tp]
\centering
\includegraphics[width=\hsize]{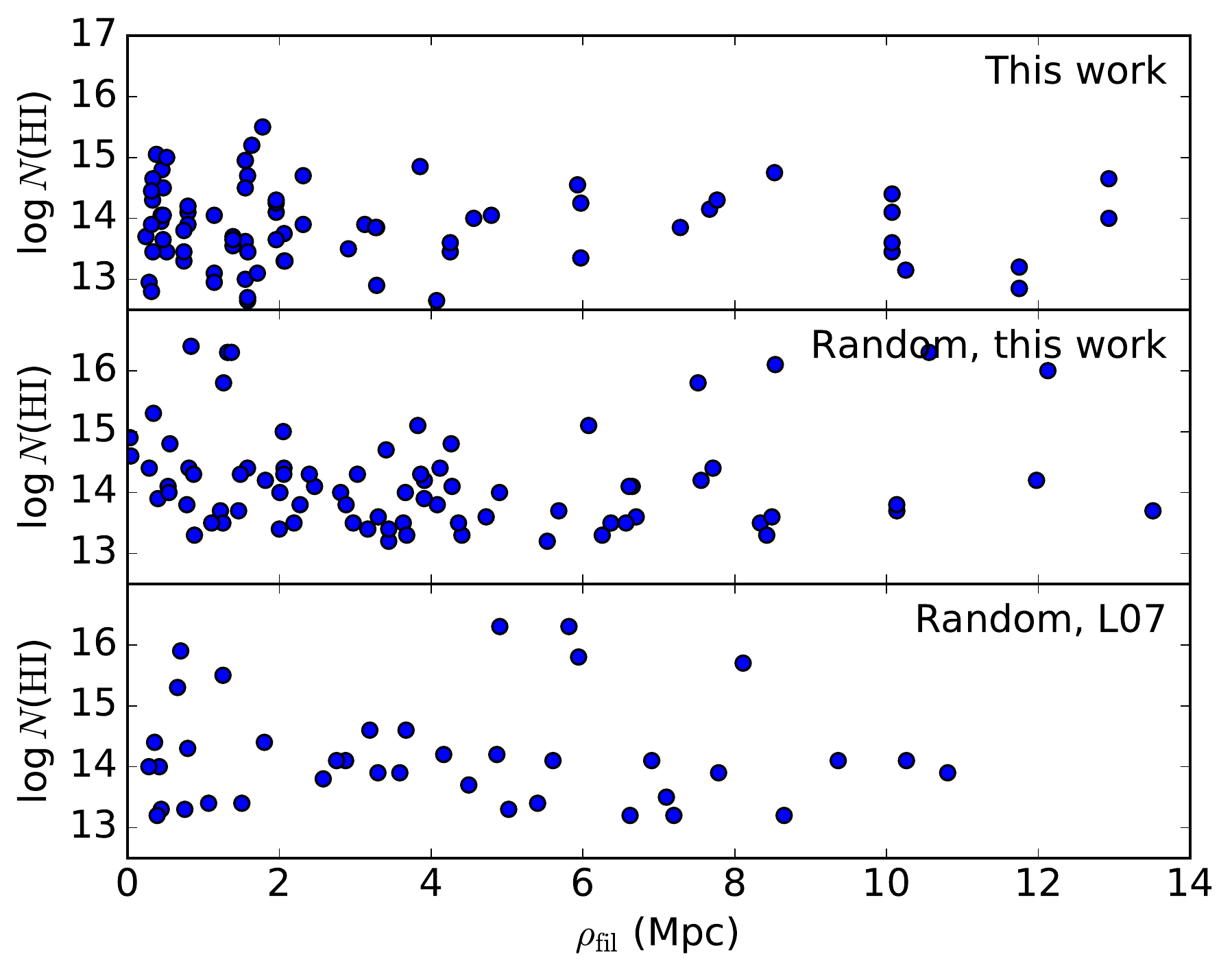}
\caption{
Same as Fig.~\ref{randomrhog}, but now for $\rho_{\rm fil}$.
}
\label{randomrhof}
\end{figure}

%__________________________________________________________________

When comparing the distance of the Ly$\alpha$ absorbers to the nearest 
filament axis, as shown in Fig.\ref{randomrhof}, a similar, albeit less 
pronounced trend can be seen. Measured absorbers are generally found 
closer to the filament axis than a random distribution shows. 
One may argue that this could be a selection effect, e.g., from targeting
particularly interesting areas such as the Virgo Cluster, which would 
result in more sightlines near the Virgo filament. 
However, Fig.\ref{EWrhof} showed the break-down of absorbers into 
different filaments, demonstrating that the absorbers belonging to 
the Virgo Cluster filament (green) are in fact more spread out than
absorbers in other regions, speaking against such a selection effect.

\begin{figure}[tp]
\centering
\includegraphics[width=\hsize]{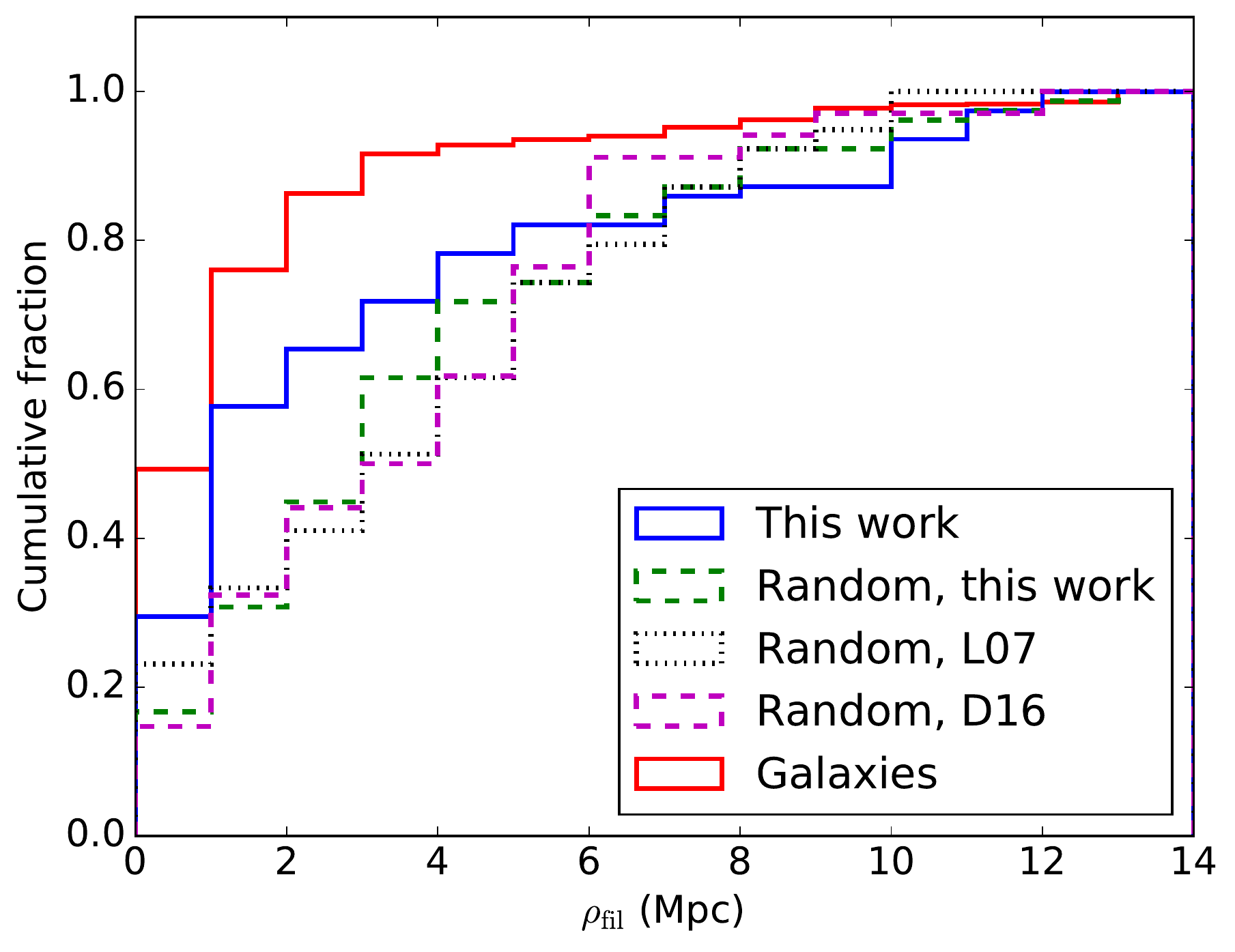}
\caption{
Cumulative distribution function for $\rho_{\rm fil}$ for three different
absorber samples. The measured absorbers in the COS data are indicated 
in blue, the random sample with own statistics is plotted in dashed 
green, the random sample with the L07 statistics is indicated in
dotted black, and the random sample from \citet{Danforth16} 
is added in dashed magenta (D16). 
The distribution for the galaxies is shown in red.
}
\label{cdistrf}
\end{figure}

To further investigate the possible clustering signal of Ly$\alpha$ 
absorbers near the filament axis, we have plotted in Figure~\ref{cdistrf} 
the cumulative distribution function for $\rho_{\rm fil}$ for the
three previously mentioned absorber samples 
(observed sample, random sample with own statistics, random sample with 
L07 statistics) as well as the galaxies that constitute the filament. 
We also have added another absorber test sample (D16)
generated from the Ly$\alpha$ column density distribution of absorbers 
reported in \citet{Danforth16}.
The cumulative distribution of galaxies set the reference point, as 
these {\it define} the filament. As can be seen, the observed 
distribution of absorbers cluster more strongly around the 
filament axis than the three random absorber test distributions, but not as 
strongly as the galaxies. Within the inner 2 Mpc, in particular, the 
fraction of measured Ly$\alpha$ absorbers rises faster than the 
synthetic absorbers in the randomised samples.

The cumulative distribution function as shown in Fig.~\ref{cdistrf}
can be compared with one for absorbers associated with galaxies. 
Fig.~3 in \citet{Penton02} shows this function for 46 Ly$\alpha$ 
absorbers and subsamples thereof. 
Their full sample follows a distribution similar to our absorbers, with $\sim$60\,\%
found within 2~Mpc of the nearest galaxy \citep{Penton02} or filament (this study).
Both studies show the galaxies more strongly clustered than the Ly$\alpha$ absorbers.
\citet{Stocke13} compared their absorber-galaxy cumulative distribution function
with a random distribution concluded that absorbers are associated with
galaxies in a more general way, i.e., tracing the large-scale structure 
instead of individual galaxies.  \citet{Penton02,Stocke13,Keeney18} all 
conclude that high-column density absorbers are more strongly correlated with
galaxies than those with lower column densities.
This is in agreement with what is found here: Ly$\alpha$ absorbers
do not trace the same distribution as galaxies, but they are not randomly
distributed around filaments either.

%__________________________________________________________________

%__________________________________________________________________

\begin{table}[bp]
\caption{Kolmogorov-Smirnov test for impact-parameter distributions}      
\label{table:KStest}    
\centering                                  
\begin{tabular}{l l l }     
\hline\hline
Sample compared    & KS statistics & $p$-value \\   
\hline                              
Random, this work  & 0.28 & 0.003 \\
Random, L07        & 0.31 & 0.011 \\
Random, D16 & 0.32 & 0.013 \\
V8k galaxies       & 0.26 & 0.032 \\
\hline                                         
\end{tabular}
\end{table}

%__________________________________________________________________

A Kolmogorov-Smirnov (KS) test confirms that the absorbers found in the COS
data are not drawn from a random distribution. Table~\ref{table:KStest} lists 
the KS values and $p$-values for the different samples. 

A KS-test can be used to compare two different samples to evaluate whether 
if they are drawn from the same parent distribution. A high
KS value (maximum of 1) indicates a high probability that this 
is true. The $p$-value instead indicates the significance that the 
null hypothesis is rejected. In this case, the null hypothesis is 
that $\rho_{\rm fil}$ for the absorbers measured in this study
follows the same distribution as that of a random sample, or of
the galaxies in the filament. A $p$-value of
$<0.05$ means the null hypothesis can be rejected with a probability of
$>95$\,\%. In all three comparisons between the measurements
and the random absorber samples (our own sample, randomised, and the 
randomised L07 and D16 samples), the KS-test indicates they do not follow
the same distribution.

The low $p$-value that we obtain from comparing the COS absorber sample 
with the V8k galaxy sample in the filaments further indicates that the 
the Ly$\alpha$ absorbers are not drawn from the same distribution as 
the galaxies. 
Evaluating the hypothesis that there are two populations of Ly$\alpha$ 
absorbers \citep[e.g.][]{Morris93, Tejos12}, we have removed from 
the sample (as a test) those absorbers that we have associated with 
galaxies. This has, however, no significant effect on the 
cumulative distribution function for $\rho_{\rm fil}$.

In conclusion, the cumulative distribution functions for $\rho_{\rm fil}$ 
show that galaxies are more strongly clustered in the filaments than
the Ly$\alpha$ absorbers that belong to the same cosmological 
structures. Ly$\alpha$ absorbers do not follow a random distribution 
and neither do they follow the same distribution as the galaxies that 
constitute large-scale filaments. There might be two (or more) separate 
populations of Ly$\alpha$ absorbers in filaments, but (from our
study) there is no evidence that the Ly$\alpha$ absorbers that are
{\it not} directly associated with large galaxies are randomly 
distributed in the field of the filament. 

Deeper insights into these aspects (including other important cosmological 
issues such as overdensity bias-factors and how they affect the absorber/galaxy/filament
statistics) are highly desirable, but will require a much larger observational
data set in combination with numerical cosmological simulations.

%%%%%%%%%%%%%%%%%%%%%%%%%%%%%%%%%%%%%%%%%%%%%%%%%%%%%%%%%%%%%%%%%%%%%%%%

\section{Summary and concluding remarks}

In this study, we have combined galaxy data of more than 30,000 nearby 
galaxies from the V8k catalogue \citet{Courtois13} with HST/COS UV 
spectral data of 302 distant AGN to investigate the relation between
intervening H\,{\sc i} Ly$\alpha$ absorbers and five nearby 
cosmological structures (galaxy filaments) at $z\approx0$ 
($v<6700$ km\,s$^{-1}$).\\
\\

(1) All in all, we identify 587 intervening Ly$\alpha$ absorbers 
along the 302 COS sightlines in the wavelength range between
$1220$ and $1243$\,\AA. For the 91 sightlines that pass the 
immediate environment of the examined galaxy filaments
we analysed in detail 215 (229) Ly$\alpha$ absorption 
systems (components) and derived column densities and $b$-values 
for H\,{\sc i} (and associated metals, if available). \\
\\
(2) For the individual galaxies in our sample, we have calculated the
virial radii from their luminosities and the galaxy impact parameters,
$\rho_{\rm gal}$, to the COS sightlines. We assume $29$ Ly$\alpha$
absorbers to be directly associated with galaxies, as they are
located with 1.5 virial radii of their host galaxies and within
400 km\,s$^{-1}$ of the galaxies' recession velocity.\\
\\
(3) We characterise the geometry of the galaxy filament by considering
the galaxy distribution in individual segments of the filaments.
In this way, we define for each filament a geometrical axis that
we use as reference for defining the filament impact parameters, 
$\rho_{\rm fil}$, for those Ly$\alpha$ absorbers that are located 
within 1000 km\,s$^{-1}$ of the filament.\\
\\
(4) We find that the absorption velocities of the Ly$\alpha$ 
absorbers reflect the large-scale velocity pattern of the four 
galaxy filaments, for which sufficient absorption-line data are 
available. 74 absorbers are aligned in position and velocity 
with the galaxy filaments, indicating that these absorbers and the 
galaxies trace the same large-scale structure .\\
\\
(5) If we relate the measured Ly$\alpha$ equivalent widths 
(or H\,{\sc i} column densities) with the galaxy and filament 
impact parameters, we find that the strongest absorbers
(equivalent widths $W_{\lambda} >500$ m\AA) are preferentially located
in the vicinity of individual galaxies (within 3 virial radii) 
and/or in the vicinity of the filament axes (within 5 Mpc).
The observed relations between $W$ and $\rho_{\rm gal}$/$\rho_{\rm fil}$ 
exhibit substantial scatter, however, disfavouring a simple
equivalent width/impact parameter anti-correlation.\\
\\
(6) We find that the measured H\,{\sc i} components 
follow a column-density distribution function with a slope
of $-\beta=-1.63\pm0.12$, a value that is typical for 
the low-redshift Ly$\alpha$ forest. Only for the 
sub-sample of absorbers within 1000~km\,s$^{-1}$ of the filament 
velocity do we obtain a shallower CDDF with $\beta = 1.47 \pm 0.24$,
possibly indicating an excess of high column-density absorbers
in galaxy filaments when compared to the overall Ly$\alpha$ forest.\\
\\
(7) The Ly$\alpha$ absorbers that lie within 1000~km\,s$^{-1}$ 
of the nearest filament have a $\sim 90$ percent higher rate of 
incidence ($d\mathcal{N}/dz=189\pm25$ for log $N$(H\,{\sc i}) $\geq 13.2$) 
than the general Ly$\alpha$ absorber population in our sample 
($d\mathcal{N}/dz=98\pm8$ for log $N$(H\,{\sc i}) $\geq 13.2$).  
This higher number density of Ly$\alpha$ absorbers per unit redshift 
most likely reflects the filaments' general matter overdensity.\\
\\
(8) We compare the filament impact-parameter distributions of 
the galaxies, measured Ly$\alpha$ absorbers, and a (synthetic) 
Ly$\alpha$ absorber sample with randomised locations on the sky
with each other. We find that 
the galaxies are most strongly clustered around the filament
axes, while the spatial clustering of the observed Ly$\alpha$ 
absorbers around the filament axes is evident, but less 
pronounced. Using a KS test, we confirm
that the Ly$\alpha$ absorbers neither follow the 
impact-parameter distribution of the galaxies, nor do they follow 
a random distribution, but represent an individual, spatially
confined sample of objects.\\
\\
Taken together, the results of our study underline that the 
relation between intervening Ly$\alpha$ absorbers, large-scale
cosmological filaments, and individual galaxies (that constitute
the filaments) in the local universe is complex and manifold, 
and difficult to reconstruct with existing data.

This complexity is not surprising, of course, if we recall, what 
Ly$\alpha$ absorbers actually are: they are objects that trace 
local gas overdensities in an extremely extended, diffuse medium 
that is gravitationally confined in hierarchically structured
potential wells, and stirred up by large-scale matter flows and
galaxy feedback. In this picture, the spatial distribution 
of Ly$\alpha$ absorbers in cosmological filaments is governed
by both the distribution of individual sinks in the large-scale
gravitational potential energy distribution (i.e., galaxies, 
galaxy groups etc.) and more (or less) stochastically distributed 
density fluctuations at larger scales that reflect the internal 
dynamics of the IGM.

For the future, we are planning to extend our study of the relation
between intervening absorbers and cosmological filaments in the
local universe by using a larger (and deeper) galaxy sample and 
additional HST/COS spectra, in combination with constrained
magneto-hydrodynamic cosmological simulations of nearby 
cosmological structures.

%%%%%%%%%%%%%%%%%%%%%%%%%%%%%%%%%%%%%%%%%%%%%%%%%%%%%%%%%%%%%%%%%%%%%%%%

\begin{acknowledgements}
The authors would like to thank the referee for his valuable 
comments and suggestions which helped to improve the manuscript.
\end{acknowledgements}

%%%%%%%%%%%%%%%%%%%%%%%%%%%%%%%%%%%%%%%%%%%%%%%%%%%%%%%%%%%%%%%%%%%%%%%%

\bibliographystyle{aa} % style aa.bst
\bibliography{Lya_in_fil}

\begin{thebibliography}{54}
\expandafter\ifx\csname natexlab\endcsname\relax\def\natexlab#1{#1}\fi

\bibitem[{{Bond} {et~al.}(2010){Bond}, {Strauss}, \& {Cen}}]{Bond10}
{Bond}, N.~A., {Strauss}, M.~A., \& {Cen}, R. 2010, \mnras, 409, 156

\bibitem[{{Bowen} {et~al.}(2002){Bowen}, {Pettini}, \& {Blades}}]{Bowen02}
{Bowen}, D.~V., {Pettini}, M., \& {Blades}, J.~C. 2002, \apj, 580, 169

\bibitem[{{Bower} {et~al.}(2006){Bower}, {Benson}, {Malbon}, {Helly}, {Frenk},
  {Baugh}, {Cole}, \& {Lacey}}]{Bower06}
{Bower}, R.~G., {Benson}, A.~J., {Malbon}, R., {et~al.} 2006, \mnras, 370, 645

\bibitem[{{Cen} \& {Ostriker}(1999)}]{Cen99}
{Cen}, R. \& {Ostriker}, J.~P. 1999, \apj, 514, 1

\bibitem[{{Chen} {et~al.}(2001){Chen}, {Lanzetta}, {Webb}, \&
  {Barcons}}]{Chen01}
{Chen}, H.-W., {Lanzetta}, K.~M., {Webb}, J.~K., \& {Barcons}, X. 2001, \apj,
  559, 654

\bibitem[{{Colless} {et~al.}(2001){Colless}, {Dalton}, {Maddox}, {Sutherland },
  {Norberg}, {Cole}, {Bland -Hawthorn}, {Bridges}, {Cannon}, {Collins},
  {Couch}, {Cross}, {Deeley}, {De Propris}, {Driver}, {Efstathiou}, {Ellis},
  {Frenk}, {Glazebrook}, {Jackson}, {Lahav}, {Lewis}, {Lumsden}, {Madgwick},
  {Peacock}, {Peterson}, {Price}, {Seaborne}, \& {Taylor}}]{Colless01}
{Colless}, M., {Dalton}, G., {Maddox}, S., {et~al.} 2001, \mnras, 328, 1039

\bibitem[{{Courtois} {et~al.}(2013){Courtois}, {Pomar{\`e}de}, {Tully},
  {Hoffman}, \& {Courtois}}]{Courtois13}
{Courtois}, H.~M., {Pomar{\`e}de}, D., {Tully}, R.~B., {Hoffman}, Y., \&
  {Courtois}, D. 2013, \aj, 146, 69

\bibitem[{{Danforth} {et~al.}(2016){Danforth}, {Keeney}, {Tilton}, {Shull},
  {Stocke}, {Stevans}, {Pieri}, {Savage}, {France}, {Syphers}, {Smith},
  {Green}, {Froning}, {Penton}, \& {Osterman}}]{Danforth16}
{Danforth}, C.~W., {Keeney}, B.~A., {Tilton}, E.~M., {et~al.} 2016, \apj, 817,
  111

\bibitem[{{Danforth} \& {Shull}(2008)}]{Danforth08}
{Danforth}, C.~W. \& {Shull}, J.~M. 2008, \apj, 679, 194

\bibitem[{{Dashtamirova} {et~al.}(2018){Dashtamirova}, {Fischer},
  {et~al.}}]{Dashtamirova19}
{Dashtamirova}, D., {Fischer}, W.~J., {et~al.} 2018, Cosmic Origins
  Spectrograph Instrument Handbook, Version 12.0 (Baltimore: STScI)

\bibitem[{{Dav{\'e}} {et~al.}(1999){Dav{\'e}}, {Hernquist}, {Katz}, \&
  {Weinberg}}]{Dave99}
{Dav{\'e}}, R., {Hernquist}, L., {Katz}, N., \& {Weinberg}, D.~H. 1999, \apj,
  511, 521

\bibitem[{{Davies} {et~al.}(2019){Davies}, {Crain}, {McCarthy}, {Oppenheimer},
  {Schaye}, {Schaller}, \& {McAlpine}}]{Davies19}
{Davies}, J.~J., {Crain}, R.~A., {McCarthy}, I.~G., {et~al.} 2019, \mnras, 485,
  3783

\bibitem[{{Erb}(2008)}]{Erb08}
{Erb}, D.~K. 2008, \apj, 674, 151

\bibitem[{{Fairall} {et~al.}(1998){Fairall}, {Woudt}, \&
  {Kraan-Korteweg}}]{Fairall98}
{Fairall}, A.~P., {Woudt}, P.~A., \& {Kraan-Korteweg}, R.~C. 1998, \aaps, 127,
  463

\bibitem[{{French} \& {Wakker}(2017)}]{French17}
{French}, D.~M. \& {Wakker}, B.~P. 2017, \apj, 837, 138

\bibitem[{{Genzel} {et~al.}(2010){Genzel}, {Tacconi}, {Gracia-Carpio},
  {Sternberg}, {Cooper}, {Shapiro}, {Bolatto}, {Bouch{\'e}}, {Bournaud},
  {Burkert}, {Combes}, {Comerford}, {Cox}, {Davis}, {Schreiber},
  {Garcia-Burillo}, {Lutz}, {Naab}, {Neri}, {Omont}, {Shapley}, \&
  {Weiner}}]{Genzel10}
{Genzel}, R., {Tacconi}, L.~J., {Gracia-Carpio}, J., {et~al.} 2010, \mnras,
  407, 2091

\bibitem[{{Green} {et~al.}(2012){Green}, {Froning}, {Osterman}, {Ebbets},
  {Heap}, {Leitherer}, {Linsky}, {Savage}, {Sembach}, {Shull}, {Siegmund},
  {Snow}, {Spencer}, {Stern}, {Stocke}, {Welsh}, {B{\'e}land}, {Burgh},
  {Danforth}, {France}, {Keeney}, {McPhate}, {Penton}, {Andrews},
  {Brownsberger}, {Morse}, \& {Wilkinson}}]{Green2012}
{Green}, J.~C., {Froning}, C.~S., {Osterman}, S., {et~al.} 2012, \apj, 744, 60

\bibitem[{{Keeney} {et~al.}(2018){Keeney}, {Stocke}, {Pratt}, {Davis},
  {Syphers}, {Danforth}, {Shull}, {Froning}, {Green}, {Penton}, \&
  {Savage}}]{Keeney18}
{Keeney}, B.~A., {Stocke}, J.~T., {Pratt}, C.~T., {et~al.} 2018, \apjs, 237, 11

\bibitem[{{Lehner} {et~al.}(2007){Lehner}, {Savage}, {Richter}, {Sembach},
  {Tripp}, \& {Wakker}}]{Lehner07}
{Lehner}, N., {Savage}, B.~D., {Richter}, P., {et~al.} 2007, \apj, 658, 680

\bibitem[{{Madau} {et~al.}(2001){Madau}, {Ferrara}, \& {Rees}}]{Madau01}
{Madau}, P., {Ferrara}, A., \& {Rees}, M.~J. 2001, \apj, 555, 92

\bibitem[{{Martizzi} {et~al.}(2019){Martizzi}, {Vogelsberger}, {Artale},
  {Haider}, {Torrey}, {Marinacci}, {Nelson}, {Pillepich}, {Weinberger},
  {Hernquist}, {Naiman}, \& {Springel}}]{Martizzi19}
{Martizzi}, D., {Vogelsberger}, M., {Artale}, M.~C., {et~al.} 2019, \mnras,
  486, 3766

\bibitem[{{Mei} {et~al.}(2007){Mei}, {Blakeslee}, {C{\^o}t{\'e}}, {Tonry},
  {West}, {Ferrarese}, {Jord{\'a}n}, {Peng}, {Anthony}, \& {Merritt}}]{Mei07}
{Mei}, S., {Blakeslee}, J.~P., {C{\^o}t{\'e}}, P., {et~al.} 2007, \apj, 655,
  144

\bibitem[{{Morris} \& {Jannuzi}(2006)}]{Morris06}
{Morris}, S.~L. \& {Jannuzi}, B.~T. 2006, \mnras, 367, 1261

\bibitem[{{Morris} {et~al.}(1993){Morris}, {Weymann}, {Dressler}, {McCarthy},
  {Smith}, {Terrile}, {Giovanelli}, \& {Irwin}}]{Morris93}
{Morris}, S.~L., {Weymann}, R.~J., {Dressler}, A., {et~al.} 1993, \apj, 419,
  524

\bibitem[{{Morton}(2003)}]{Morton03}
{Morton}, D.~C. 2003, \apjs, 149, 205

\bibitem[{{Pallottini} {et~al.}(2014){Pallottini}, {Gallerani}, \&
  {Ferrara}}]{Pallottini14}
{Pallottini}, A., {Gallerani}, S., \& {Ferrara}, A. 2014, \mnras, 444, L105

\bibitem[{{Penton} {et~al.}(2000){Penton}, {Shull}, \& {Stocke}}]{Penton00}
{Penton}, S.~V., {Shull}, J.~M., \& {Stocke}, J.~T. 2000, \apj, 544, 150

\bibitem[{{Penton} {et~al.}(2002){Penton}, {Stocke}, \& {Shull}}]{Penton02}
{Penton}, S.~V., {Stocke}, J.~T., \& {Shull}, J.~M. 2002, \apj, 565, 720

\bibitem[{{Prochaska} {et~al.}(2011){Prochaska}, {Weiner}, {Chen}, {Mulchaey},
  \& {Cooksey}}]{Prochaska11}
{Prochaska}, J.~X., {Weiner}, B., {Chen}, H.~W., {Mulchaey}, J., \& {Cooksey},
  K. 2011, \apj, 740, 91

\bibitem[{{Prochaska} \& {Wolfe}(2009)}]{Prochaska09}
{Prochaska}, J.~X. \& {Wolfe}, A.~M. 2009, \apj, 696, 1543

\bibitem[{{Richter} {et~al.}(2006{\natexlab{a}}){Richter}, {Fang}, \&
  {Bryan}}]{Richter06}
{Richter}, P., {Fang}, T., \& {Bryan}, G.~L. 2006{\natexlab{a}}, \aap, 451, 767

\bibitem[{{Richter} {et~al.}(2013){Richter}, {Fox}, {Wakker}, {Lehner}, {Howk},
  {Bland-Hawthorn}, {Ben Bekhti}, \& {Fechner}}]{Richter13}
{Richter}, P., {Fox}, A.~J., {Wakker}, B.~P., {et~al.} 2013, \apj, 772, 111

\bibitem[{{Richter} {et~al.}(2011){Richter}, {Krause}, {Fechner}, {Charlton},
  \& {Murphy}}]{Richter11}
{Richter}, P., {Krause}, F., {Fechner}, C., {Charlton}, J.~C., \& {Murphy},
  M.~T. 2011, \aap, 528, A12

\bibitem[{{Richter} {et~al.}(2017){Richter}, {Nuza}, {Fox}, {Wakker}, {Lehner},
  {Ben Bekhti}, {Fechner}, {Wendt}, {Howk}, {Muzahid}, {Ganguly}, \&
  {Charlton}}]{Richter17}
{Richter}, P., {Nuza}, S.~E., {Fox}, A.~J., {et~al.} 2017, \aap, 607, A48

\bibitem[{{Richter} {et~al.}(2008){Richter}, {Paerels}, \&
  {Kaastra}}]{Richter08}
{Richter}, P., {Paerels}, F.~B.~S., \& {Kaastra}, J.~S. 2008, \ssr, 134, 25

\bibitem[{{Richter} {et~al.}(2006{\natexlab{b}}){Richter}, {Savage}, {Sembach},
  \& {Tripp}}]{Richter06b}
{Richter}, P., {Savage}, B.~D., {Sembach}, K.~R., \& {Tripp}, T.~M.
  2006{\natexlab{b}}, \aap, 445, 827

\bibitem[{{Richter} {et~al.}(2016){Richter}, {Wakker}, {Fechner}, {Herenz},
  {Tepper-Garc{\'\i}a}, \& {Fox}}]{Richter16}
{Richter}, P., {Wakker}, B.~P., {Fechner}, C., {et~al.} 2016, \aap, 590, A68

\bibitem[{{Saunders} {et~al.}(2000{\natexlab{a}}){Saunders}, {d'Mellow},
  {Tully}, {Mobasher}, {Carrasco}, {Maddox}, {Hau}, {Sutherland}, {Clements},
  \& {Staveley-Smith}}]{Saunders00a}
{Saunders}, W., {d'Mellow}, K.~J., {Tully}, R.~B., {et~al.} 2000{\natexlab{a}},
  in Astronomical Society of the Pacific Conference Series, Vol. 218, Mapping
  the Hidden Universe: The Universe behind the Mily Way - The Universe in HI,
  ed. R.~C. {Kraan-Korteweg}, P.~A. {Henning}, \& H.~{Andernach}, 153

\bibitem[{{Saunders} {et~al.}(2000{\natexlab{b}}){Saunders}, {Sutherland},
  {Maddox}, {Keeble}, {Oliver}, {Rowan-Robinson}, {McMahon}, {Efstathiou},
  {Tadros}, {White}, {Frenk}, {Carrami{\~n}ana}, \& {Hawkins}}]{Saunders00b}
{Saunders}, W., {Sutherland}, W.~J., {Maddox}, S.~J., {et~al.}
  2000{\natexlab{b}}, \mnras, 317, 55

\bibitem[{{Shaya} {et~al.}(1995){Shaya}, {Peebles}, \& {Tully}}]{shaya95}
{Shaya}, E.~J., {Peebles}, P.~J.~E., \& {Tully}, R.~B. 1995, \apj, 454, 15

\bibitem[{{Shull}(2003)}]{Shull03}
{Shull}, J.~M. 2003, in Astrophysics and Space Science Library, Vol. 281, The
  IGM/Galaxy Connection. The Distribution of Baryons at z=0, ed. J.~L.
  {Rosenberg} \& M.~E. {Putman}, 1

\bibitem[{{Shull}(2014)}]{Shull14}
{Shull}, J.~M. 2014, \apj, 784, 142

\bibitem[{{Shull} {et~al.}(2012){Shull}, {Smith}, \& {Danforth}}]{Shull12}
{Shull}, J.~M., {Smith}, B.~D., \& {Danforth}, C.~W. 2012, \apj, 759, 23

\bibitem[{{Stocke} {et~al.}(2013){Stocke}, {Keeney}, {Danforth}, {Shull},
  {Froning}, {Green}, {Penton}, \& {Savage}}]{Stocke13}
{Stocke}, J.~T., {Keeney}, B.~A., {Danforth}, C.~W., {et~al.} 2013, \apj, 763,
  148

\bibitem[{{Tejos} {et~al.}(2012){Tejos}, {Morris}, {Crighton}, {Theuns},
  {Altay}, \& {Finn}}]{Tejos12}
{Tejos}, N., {Morris}, S.~L., {Crighton}, N. H.~M., {et~al.} 2012, \mnras, 425,
  245

\bibitem[{{Tejos} {et~al.}(2014){Tejos}, {Morris}, {Finn}, {Crighton},
  {Bechtold}, {Jannuzi}, {Schaye}, {Theuns}, {Altay}, {Le F{\`e}vre},
  {Ryan-Weber}, \& {Dav{\'e}}}]{Tejos14}
{Tejos}, N., {Morris}, S.~L., {Finn}, C.~W., {et~al.} 2014, \mnras, 437, 2017

\bibitem[{{Tejos} {et~al.}(2016){Tejos}, {Prochaska}, {Crighton}, {Morris},
  {Werk}, {Theuns}, {Padilla}, {Bielby}, \& {Finn}}]{Tejos16}
{Tejos}, N., {Prochaska}, J.~X., {Crighton}, N. H.~M., {et~al.} 2016, \mnras,
  455, 2662

\bibitem[{{Tilton} {et~al.}(2012){Tilton}, {Danforth}, {Shull}, \&
  {Ross}}]{Tilton12}
{Tilton}, E.~M., {Danforth}, C.~W., {Shull}, J.~M., \& {Ross}, T.~L. 2012,
  \apj, 759, 112

\bibitem[{{Tully} {et~al.}(2009){Tully}, {Rizzi}, {Shaya}, {Courtois},
  {Makarov}, \& {Jacobs}}]{Tully09}
{Tully}, R.~B., {Rizzi}, L., {Shaya}, E.~J., {et~al.} 2009, arXiv e-prints,
  arXiv:0902.3668

\bibitem[{{Tumlinson} {et~al.}(2002){Tumlinson}, {Shull}, {Rachford},
  {Browning}, {Snow}, {Fullerton}, {Jenkins}, {Savage}, {Crowther}, {Moos},
  {Sembach}, {Sonneborn}, \& {York}}]{Tumlinson02}
{Tumlinson}, J., {Shull}, J.~M., {Rachford}, B.~L., {et~al.} 2002, \apj, 566,
  857

\bibitem[{{Tytler}(1987)}]{Tytler87}
{Tytler}, D. 1987, \apj, 321, 49

\bibitem[{{Wakker} {et~al.}(2015){Wakker}, {Hernandez}, {French}, {Kim},
  {Oppenheimer}, \& {Savage}}]{Wakker15}
{Wakker}, B.~P., {Hernandez}, A.~K., {French}, D.~M., {et~al.} 2015, \apj, 814,
  40

\bibitem[{{Wakker} \& {Savage}(2009)}]{Wakker09}
{Wakker}, B.~P. \& {Savage}, B.~D. 2009, \apjs, 182, 378

\bibitem[{{York} {et~al.}(2000){York}, {Adelman}, {Anderson}, {Anderson},
  {Annis}, {Bahcall}, {Bakken}, {Barkhouser}, {Bastian}, {Berman}, {Boroski},
  {Bracker}, {Briegel}, {Briggs}, {Brinkmann}, {Brunner}, {Burles}, {Carey},
  {Carr}, {Castander}, {Chen}, {Colestock}, {Connolly}, {Crocker}, {Csabai},
  {Czarapata}, {Davis}, {Doi}, {Dombeck}, {Eisenstein}, {Ellman}, {Elms},
  {Evans}, {Fan}, {Federwitz}, {Fiscelli}, {Friedman}, {Frieman}, {Fukugita},
  {Gillespie}, {Gunn}, {Gurbani}, {de Haas}, {Haldeman}, {Harris}, {Hayes},
  {Heckman}, {Hennessy}, {Hindsley}, {Holm}, {Holmgren}, {Huang}, {Hull},
  {Husby}, {Ichikawa}, {Ichikawa}, {Ivezi{\'c}}, {Kent}, {Kim}, {Kinney},
  {Klaene}, {Kleinman}, {Kleinman}, {Knapp}, {Korienek}, {Kron}, {Kunszt},
  {Lamb}, {Lee}, {Leger}, {Limmongkol}, {Lindenmeyer}, {Long}, {Loomis},
  {Loveday}, {Lucinio}, {Lupton}, {MacKinnon}, {Mannery}, {Mantsch}, {Margon},
  {McGehee}, {McKay}, {Meiksin}, {Merelli}, {Monet}, {Munn}, {Narayanan},
  {Nash}, {Neilsen}, {Neswold}, {Newberg}, {Nichol}, {Nicinski}, {Nonino},
  {Okada}, {Okamura}, {Ostriker}, {Owen}, {Pauls}, {Peoples}, {Peterson},
  {Petravick}, {Pier}, {Pope}, {Pordes}, {Prosapio}, {Rechenmacher}, {Quinn},
  {Richards}, {Richmond}, {Rivetta}, {Rockosi}, {Ruthmansdorfer}, {Sand ford},
  {Schlegel}, {Schneider}, {Sekiguchi}, {Sergey}, {Shimasaku}, {Siegmund},
  {Smee}, {Smith}, {Snedden}, {Stone}, {Stoughton}, {Strauss}, {Stubbs},
  {SubbaRao}, {Szalay}, {Szapudi}, {Szokoly}, {Thakar}, {Tremonti}, {Tucker},
  {Uomoto}, {Vanden Berk}, {Vogeley}, {Waddell}, {Wang}, {Watanabe},
  {Weinberg}, {Yanny}, {Yasuda}, \& {SDSS Collaboration}}]{York00}
{York}, D.~G., {Adelman}, J., {Anderson}, John~E., J., {et~al.} 2000, \aj, 120,
  1579

\end{thebibliography}

%%%%%%%%%%%%%%%%%%%%%%%%%%%%%%%%%%%%%%%%%%%%%%%%%%%%%%%%%%%%%%%%%%%%%%%%

\clearpage
\begin{appendix}
\onecolumn
\section{Signal-to-noise}

%__________________________________________________________________

\begin{figure*}[hbt!]
\centering
\includegraphics[width=0.8\textwidth]{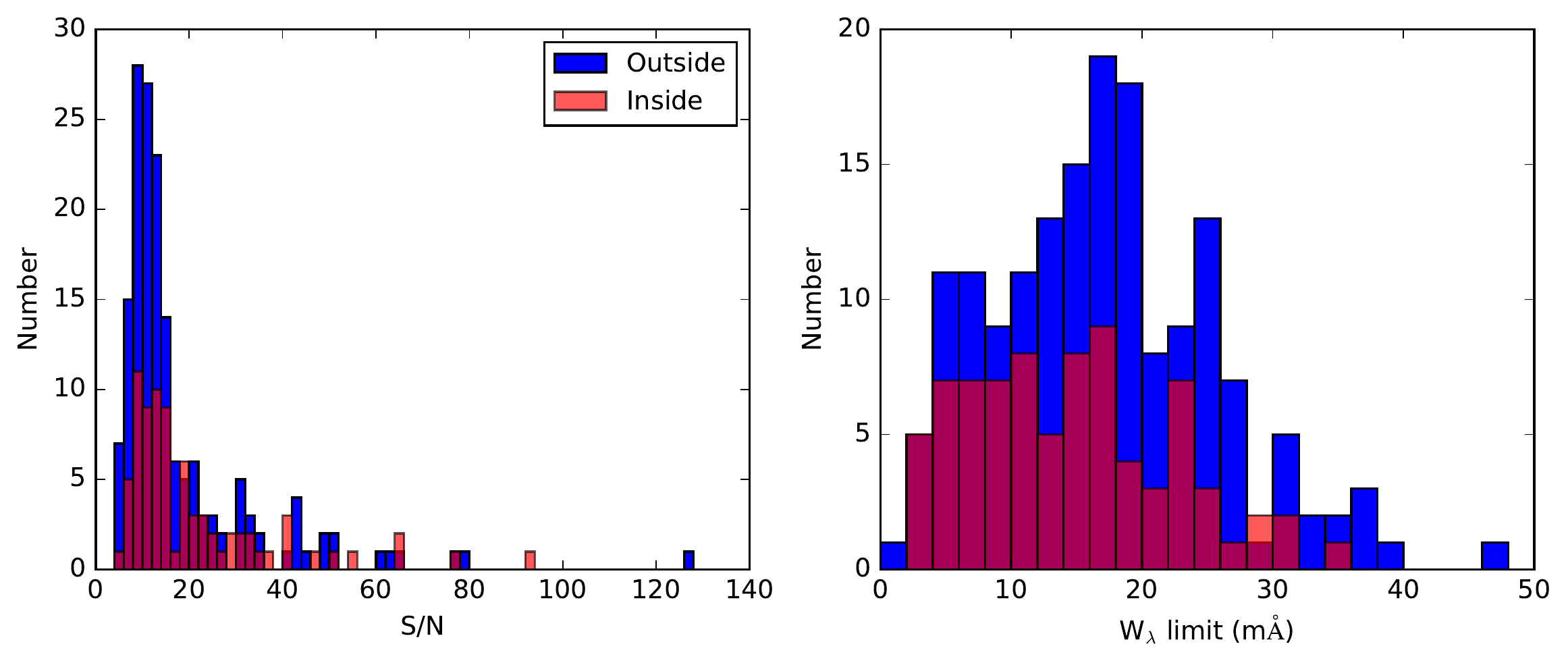}
\caption{ {\it Left panel:}
distribution of the measured S/N ratios per resolution element near \AA\,1240 
for the COS spectra that are filament-related and those outside of filaments.
{\it Right panel:} formal 3$\sigma$ detection limits for H\,{\sc i} Ly\,$\alpha$ absorption
in these spectra, based on the equation given in \citet[]{Tumlinson02}. Note that these
values reflect the detectability of Ly\,$\alpha$ absorption as a function of the 
local S/N under idealised conditions (no blending, no fixed-pattern noise, 
perfectly known continuum flux).
}
\label{SN}
\end{figure*}

%__________________________________________________________________

\section{Absorbers associated with galaxies}

%__________________________________________________________________

\begin{figure*}[hbt!]
\centering
\includegraphics[width=0.9\textwidth]{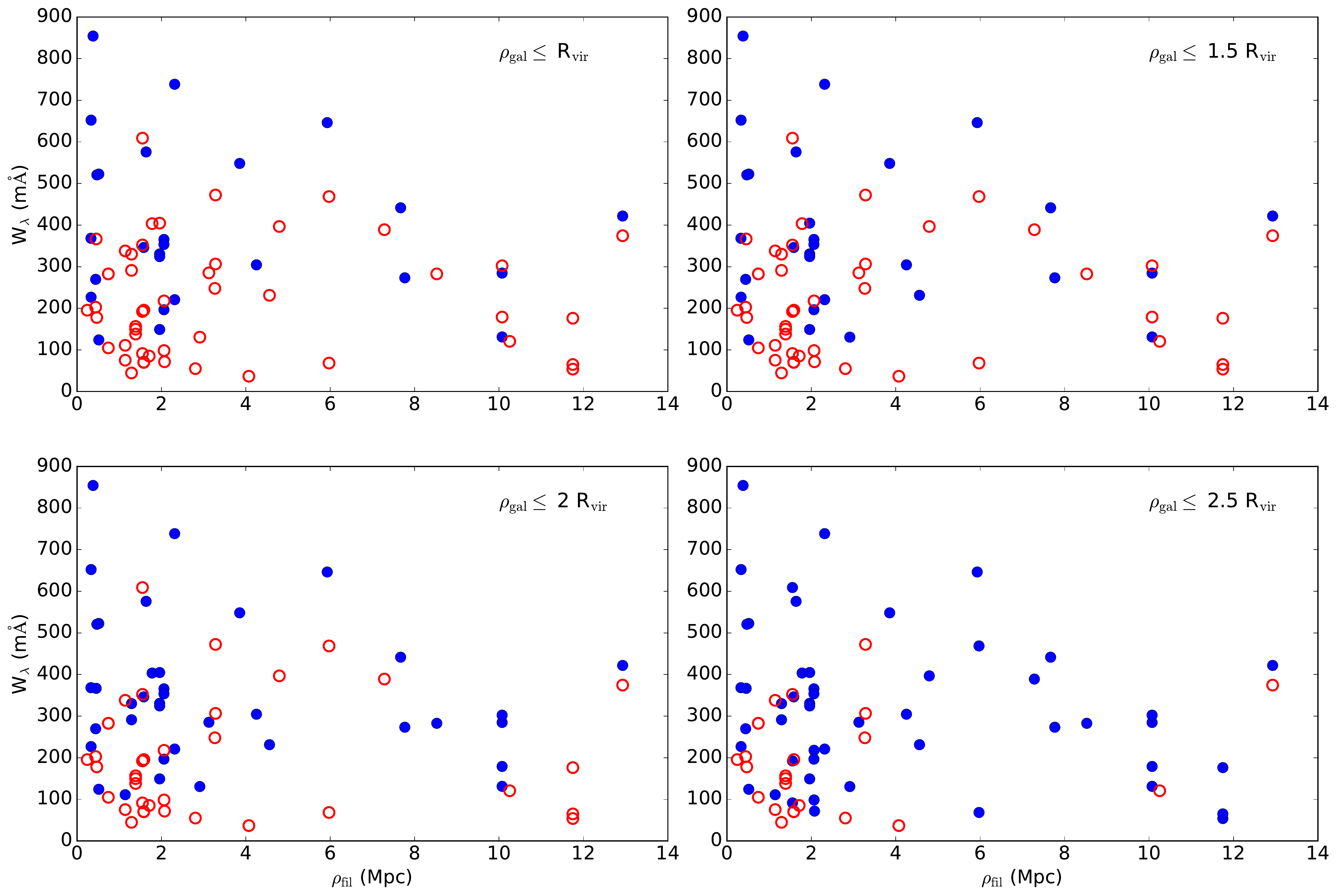}
\caption{
Same as Fig.~\ref{EWrhof}, but now using different impact-parameter criteria $\rho_{gal}$ for absorbers to be 
associated with a galaxy. Blue, filled dots are associated with a galaxy,
red, open dots are not.
}
\label{diffrhog}
\end{figure*}

%__________________________________________________________________

\end{appendix}
\end{document}